\documentclass[letterpaper,12pt]{article}
\pdfoutput=1

\usepackage{graphicx} 
\usepackage{jheppub}
\usepackage{hyperref}
\usepackage{cancel, amsmath}
\usepackage{amsfonts}
\usepackage{graphicx}
\usepackage{caption}
\usepackage{subcaption}
\usepackage{placeins}
\usepackage{tikz}
\usepackage{tikz-feynman}
\usepackage{feynmp-auto}

\usepackage{algorithm,algpseudocode}

\usepackage[english]{babel}
\usepackage{amsthm}

\newtheorem{example}{Example}

\theoremstyle{remark}
\newtheorem{remark*}{Remark}

\usepackage{array}
\usepackage{mathtools}
\usepackage{amsmath}
\usepackage{amsfonts}

\newcommand{\symU}{\ensuremath{\mathcal{U}}}
\newcommand{\symUtr}{\ensuremath{\mathcal{U}^{\mathrm{tr}}}}
\newcommand{\symF}{\ensuremath{\mathcal{F}}}
\newcommand{\symFtr}{\ensuremath{\mathcal{F}^{\mathrm{tr}}}}

\title{
All Loop Scattering As A Sampling Problem
} 

\author[a,b]{Giulio Salvatori}

\affiliation[a]{School of Natural Sciences, Institute for Advanced Study, Princeton, NJ, 08540, USA}
\affiliation[b]{Max-Plank-Instit\"ut fur Physik, Werner-Heisenberg-Institut, D–85748 Garching bei M\"unchen, Germany}
\emailAdd{salvatori@ias.edu}

\date{\today}

\abstract{
How to turn the flip of a coin into a random variable whose expected value equals a scattering amplitude?
We answer this question by constructing a numerical algorithm to evaluate curve integrals --- a novel formulation of scattering amplitudes --- by a Monte Carlo strategy.
To achieve a satisfactory accuracy we take advantage of tropical importance sampling.
The crucial result is that the sampling procedure can be realized as a \emph{stochastic process on surfaces} which can be simulated efficiently on a computer.
The key insight is to let go of the Feynman-bias that amplitudes should be presented as a sum over diagrams, 
and instead re-arrange the sum as suggested by a dual decomposition of curve integrals.
We attach an implementation of this algorithm as an ancillary file, which we have used to evaluate amplitudes for the massive $\mathrm{Tr}(\phi)^3$ theory in $D=2$ space-time dimensions, up to 10-loops.
Interestingly, we observe experimentally that the number of sample points required to achieve a fixed accuracy remains significantly smaller than what the number of diagrams would suggest.
Finally we propose an extension of our method which is inspired by ideas from artificial intelligence. We use the stochastic process to define a parametrization for a space of distributions, where we formulate importance sampling for an arbitrary curve integrand as a convex optimization problem.
} 

\begin{document}

\maketitle

\section{Introduction}

\begin{figure}
    \centering
    \includegraphics[width=.75\linewidth]{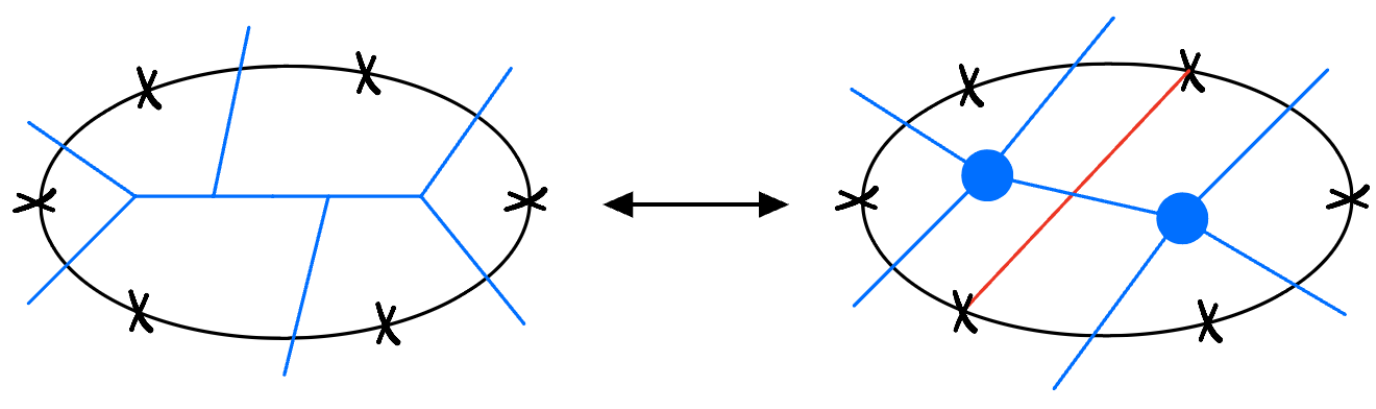}
    \caption{\emph{The basic duality of surfaceology.} Feynman graphs are replaced by curves which cuts through their propagators.}
    \label{fig:graphcurve}
\end{figure}

Surfaceology is a novel formulation for scattering amplitudes which was recently introduced in \cite{counting1,counting2}.
At its center is a simple duality that shifts the attention from the set of diagrams that can be drawn on a surface to the set of \emph{curves}, see Figure \ref{fig:graphcurve}.
This leads to a new formula for scattering amplitudes:
rather than summing over many Feynman diagrams, we compute a \emph{single integral}.
It is built out of curves and thus gets the name of curve integral.

The sharpest indication that this is not a mere repackaging of many diagrams under a common integral sign is the simplicity of the curve integrand: it can be evaluated at any point of the integration domain in an amount of time that scales polynomially with the number of particles.
This is striking since the number of diagrams contributing to the amplitude scales much more poorly. Even in the case of gauge theories, which are simpler thanks to color decomposition, the number of diagrams grows exponentially with the number of particles.

Originally, surfaceology focused on amplitudes in the $\mathrm{Tr\ } \phi^3$ theory. Surprisingly, the new formulation revealed how this seemingly simple theory contains in it the basic structure describing a much wider class of theories.
By now, it is well understood how to formulate amplitudes in arbitrary scalar colored theories \cite{tropicalscalars} and how to include fermions \cite{De:2024wsy}. Even theories describing the real world, such as pure Yang-Mills theory, have been partially brought into the fold \cite{scaffolding}.
Furthermore, the formalism inevitably contains colorless particles, and thus seems very plausible that eventually gravitational theories will be included as well.

Surfaceology is still in its infancy, but it has already unearthed many treasures in scattering amplitudes.
On one hand, it allows us to calculate loop integrands recursively, circumventing the issues that plague the extension of BCFW-like formulae \cite{Britto_2005} beyond the planar limit of supersymmetric theories \cite{Salvatori_2019, theymintegrand, cutequation}.
On the other hand, curve integrals have exposed novel properties of scattering amplitudes such as hidden zeroes, patterns of factorizations, and relations between different theories \cite{NLSM1,NLSM2,NLSM3,tropicalscalars,scaffolding}.

But, perhaps, the boldest idea about what to do with curve integrals remains the simplest one: \emph{Numerical Integration}.
In this paper we begin exploring this aspect from a Monte Carlo point of view.

A Monte Carlo strategy is well suited to our setting because the dimensionality of curve integrals grows mildly: linearly with respect to both number of particles and loops.
The method consists in sampling the domain of integration, and using the average of the integrand on the sampled points as an estimator for the true value of the integral.
The quality of the estimator depends crucially on the choice of sampling distribution. Especially in higher dimensions, the integrand may be concentrated in surprising ways around corners of the domain of integration. If these are sampled only sporadically the estimator will have a poor accuracy.
There are several general purpose techniques, loosely referred to as importance sampling, that have been thought of to address this issue.
However, in the context of curve integrals, we expect that the best results will be obtained starting from a sampling strategy that takes into account the specific geometrical structures underlying them.

In this spirit, a breakthrough was recently achieved in the context of Feynman integrals. In \cite{Borinsky:2020rqs} it was proposed to use the \emph{tropicalization} of the Feynman integrand as a sampling distribution. The motivation is that this tropical limit captures the leading term of a series expansion of the integrand around various corners of the integration space.
Furthermore, an efficient implementation of this idea was found for those diagrams that are described by  a class of polytopes called generalized permutohedra \cite{postnikov2007facesgeneralizedpermutohedra}. 
The paper even ventured in hoping that a similar strategy could be used to evaluate entire amplitudes, and not just individual diagrams, given the growing evidence that ``superior structures triangulated by Feynman diagrams'' exist.
In this paper we will see how numerical surfaceology concretely achieves this dream.

This is a good place to make an important distinction. Surfaceology is based on two pillars\footnote{Or ``miracles'', as they were called in \cite{counting1}}. The first is the existence of a geometrical space triangulated by Feynman diagrams, the \emph{Feynman polyhedral fan}.
The second is the existence of a piecewise parametrization of the Feynman fan that anchors curve integrals to a deeper structure, an elementary counting problem associated to curves on surfaces.

These two layers of the story manifest themselves in a peculiar feature of curve integrals, that of having a piecewise structure. A curve integrand is built out of expressions that are polynomials in certain piecewise linear functions, the \emph{headlight functions}, which play the role of familiar Schwinger parameters. 
It is the functional form of these headlight functions that defines the parametrization of the Feynman fan.
In this paper, however, we will think of the headlight functions as \emph{variables} and tropicalize the curve integrand with respect to them. 
The result of this procedure is a new curve integrand, $\mathcal{I}^{\rm tr}$, which in every Feynman simplex restricts to the tropical integrand of \cite{Borinsky:2020rqs}, and thus captures globally the leading behaviour of the physical integrand. 
The central problem addressed in this paper is how to efficiently generate sample points on the Feynman fan, distributed according to $\mathcal{I}^{\rm tr}$.

Somewhat disappointingly, then, the second miracle does not play any role here. 
This could be either the clue that a better sampling algorithm exists, a point on which we will comment further in the conclusions, or that the purpose in life of this additional structure is a different one, perhaps that of providing us with the \emph{best} integrand to integrate \cite{NLSM1, NLSM2, NLSM3, theymintegrand}.
Beyond these conjectures, the parametrization of the fan immediately allows to uplift our algorithm to sample over the moduli space of Riemann surfaces, opening a new door on the exploration of numerical methods to evaluate string scattering amplitudes (and generalizations thereof). For recent progress on this topic, see
\cite{Eberhardt:2023xck,Baccianti:2025gll,Banerjee:2024ibt,Eberhardt:2024twy,Arkani-Hamed:2024nzc}.
Therefore, simply by restoring the dependence of the headlight functions, we expect that the results presented here will play a key role beyond the immediate scope of this paper.

Having clarified this point, let us now turn to the problem of sampling according to $\mathcal{I}^{\rm tr}$.
The first order of business is calculating the \emph{partition function} of the distribution, that is the curve integral associated to $\mathcal{I}^{\rm tr}$:
any formula for the partition function automatically comes together with an algorithm to generate samples. 

The triangulation of the fan in Feynman simplices gives a formula for the partition function as the sum of the Hepp bounds \cite{Panzer:2019yxl,Borinsky:2020rqs} of the graphs that live on the surface (for this reason, we will also refer to the partition function as the surface Hepp bound). 
This formula corresponds to the most direct sampling algorithm.
First enumerate all the Feynman diagrams and compute their Hepp bounds.
Then randomly draw from this list, picking a diagram with a probability equal to its fractional contribution to the total surface Hepp bound.
Finally, choose any of the simplices in the Feynman fan that are labeled by this diagram, and use the algorithm described in \cite{Borinsky:2020rqs} to sample a point in it.

Evidently this is not a satisfactory algorithm. The reason is that breaking so crudely the curve integrals back into Feynman diagrams forgets entirely about any sort of ``magic'' that is spread among several of them.
And in practice this algorithm simply becomes \emph{unfeasible} very quickly due to the large number of diagrams that must be enumerated and stored in the memory of a computer.

\begin{figure}[h!]
    \centering
\includegraphics[width=.6\linewidth]{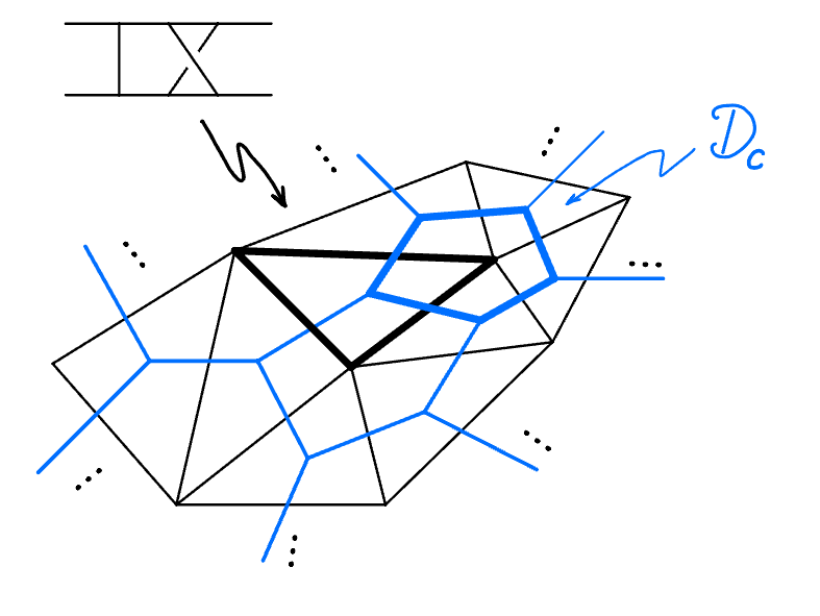}
    \centering 
    \caption{\emph{The dual barycentric decomposition of the Feynman fan.} The black triangles are Feynman simplices. The blue cells, labeled by curves, form the dual.}
    \label{fig:barycentric}
\end{figure}

We will rectify the situation by preserving some of the structure of curve integrals.
To do so, we consider a different decomposition of the domain of integration. It is the dual barycentric decomposition of the Feynman triangulation, shown in Figure \ref{fig:barycentric}. 
Integrating over the dual cells $\mathcal{D}_C$ results in a \emph{recursive formula} for the partition function, which can be calculated in terms of simpler ones associated to scattering processes with fewer particles and/or loops.

\captionsetup[figure]{name=\emph{Figure}}

\begin{figure}
    \centering
    \includegraphics[angle=-90, width=\linewidth]{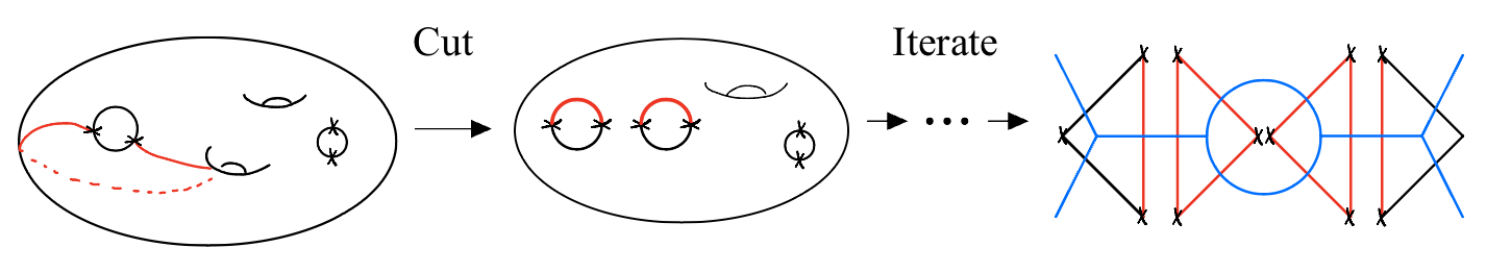}
    \caption{\emph{The stochastic process on surfaces.} At each step a curve is randomly drawn on a surface and then cut to produce a surface for the next step.}
    \label{fig:stocproc}
\end{figure}

The new formula for the partition function gives an alternative sampling algorithm.
It is based on a \emph{stochastic process on surfaces}. Rather than drawing once from the set of diagrams on a surface, we draw multiple times from the set of curves on surfaces. At each step we cut along the curve that has been drawn, producing a simpler surface from which we draw a new curve at the next step.
After a number of curves have been drawn\footnote{Equal to the dimension of the domain of integration.}, the surface has been cut down to a collection of triangles, glued by propagators flowing through their sides. This is a representation of a Feynman (fat)graph, see Figure \ref{fig:stocproc}. 
The order in which curves have been drawn during the process induces an ordering of the edges of the graph. It corresponds to a region (or ``sector'') in the Feynman simplex from which we can finally sample uniformly (in appropriate logarithmic variables) a point.

\begin{figure}
    \centering
    \includegraphics[width=0.35\linewidth,angle=-90]{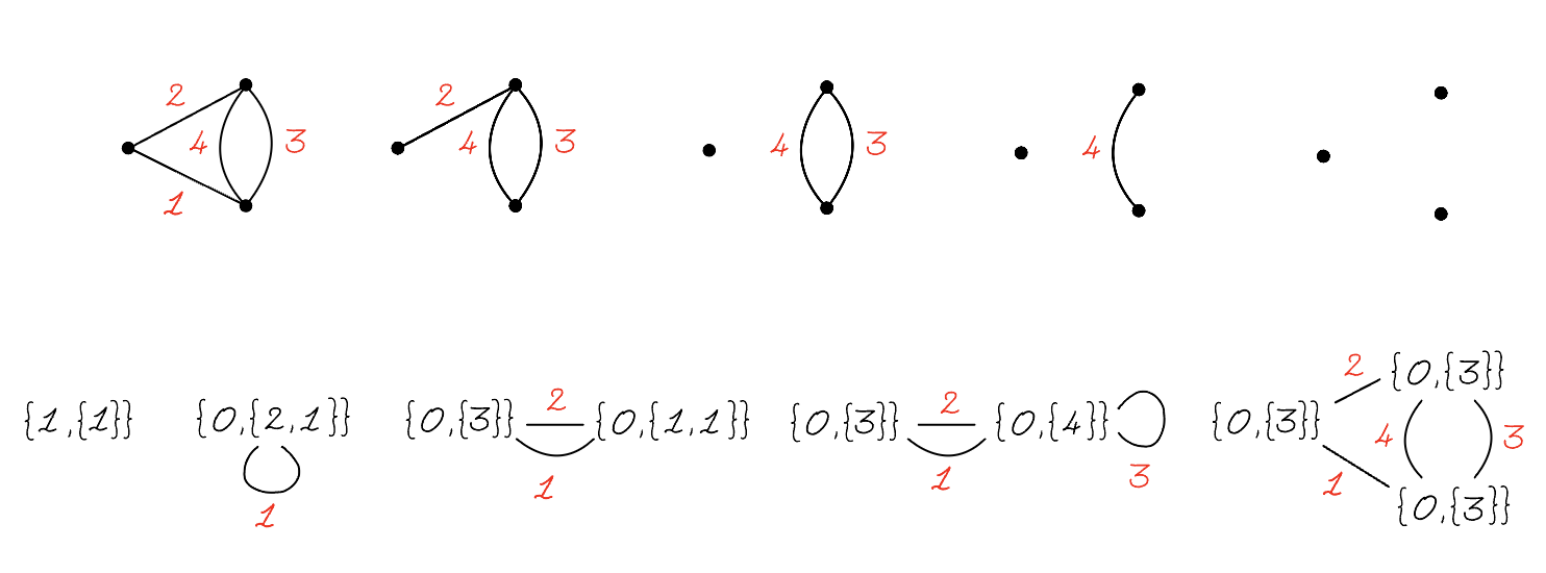}
    \caption{\emph{Dual processes.} 
    The stochastic process described in \cite{Borinsky:2020rqs} starts from a graph and \emph{removes} its edges (Top).
    The dual process introduced here starts from a vertex labeled by a surface and \emph{adds} edges (Bottom). The combinatorial labeling for surfaces used in the picture will be explained later.}
    \label{fig:dualproc1}
\end{figure}

The stochastic process on surfaces is described by the same decision tree of the direct Feynman algorithm described earlier, that prepends the sampling of a diagram by its Hepp bound to Algorithm 4 of \cite{Borinsky:2020rqs}. However, the two processes travel in the decision tree in \emph{opposite directions}, as illustrated in Figure \ref{fig:dualproc1}. Accordingly, we call ``dual'' the new sampling algorithm presented here.

The advantage is subtle, but decisive.
The dual curve sampling algorithm remains feasible long after the direct Feynman algorithm has ceased to be.
This is essentially because the dual algorithm does not need to explicitly enumerate diagrams, but only curves on surfaces, of which there are substantially fewer. 

In the rest of this paper we will illustrate this idea in detail. 
We provide a \verb|Mathematica| notebook implementing the dual sampling algorithm. It serves both as a proof-of-principle, and as an illustrative complement to the paper for readers interested in algorithms.  
Currently only the evaluation of amplitudes at zero momentum is implemented, but an extension for generic momenta, or for general theories, is straightforward.
We have used the provided code to evaluate tadpole amplitudes up to ten loops, running it in parallel on a computing cluster and we report the results in Table \ref{tab:results}.

The most interesting result concerns the number of sample points necessary to achieve a fixed accuracy. As the loop order increases, we find that it grows significantly less than the number of diagrams would naively suggest.
If we represent $A_S$ as a sum over fatgraphs, and we observe that it is necessary to generate at least $n_\Gamma$ samples to evaluate a typical Feynman diagram with $1\%$ accuracy, then we could conclude that in order to estimate $A_S$ to $1\%$ accuracy it must be necessary to generate around $n_\Gamma \times |\rm{FatGraphs}(S)|$ samples in total.
This conclusion, however, is biased by the assumption that $A_S$ should be presented as a sum over fatgraphs. We may find a different representation requiring fewer samples in total. 
Let us illustrate this point with a simple example. Suppose that one is interested in estimating the value of $\int_0^1 x\ dx$ by a Monte Carlo integration.
Someone could show up with a cumbersome formula for this integral, obtained by decomposing the domain of integration into millions of smaller segments. However, this should not discourage us into believing that we will need to generate millions of samples in order to estimate the integral of such a simple smooth function!
Similarly, the vast difference between the required number of samples observed and the one expected by the Feynman expansion, suggests that the physical integrand is approaching a ``smooth'' distribution over the Feynman fan; one for which the expansion in diagrams gives an increasingly ineffective description.

This paper focuses only on a very restricted setting: $\mathrm{Tr\ } \phi^3$ theory, in $D=2$ space-time dimensions and with internal masses for all propagators\footnote{This makes so that the amplitudes are finite.}. 
However, it is natural to extend the basic idea to more general  theories.
In order to do so, we would first have to find an appropriate \emph{parametric} curve integrand. Furthermore, in the presence of ultraviolet or infrared divergences these would have to be removed directly at the level of the integrand. If the final integrand can be evaluated quickly at a random integration point, then we are in a good position to start integrating it with the dual sampling algorithm.

The last obstacle is to understand the tropicalization of the resulting curve integrand: a problem which is made especially hard by the proliferation of minus signs generated by the process of renormalization.
We propose an alternative to tropical sampling which is inspired by ideas from artificial intelligence.  
We think of the stochastic process as \emph{defining} a class of distributions from which we can sample efficiently by construction.
We then explore this space of distributions, looking for the optimal choice to numerically integrate a desired physical integrand.
This can be formulated as an optimization problem for a suitable cost function.
Assuming the existence of an Euclidean region for a target physical integrand, we find natural to consider the Kullback–Leibler divergence, since it restricts to a convex cost function on the parametric space of distributions. 
The optimization problem can therefore be approached by gradient descent.

This idea results in a concrete strategy to evaluate any amplitude for which a positive and finite curve integrand is available in parametric form, regardless of whether its tropicalization is under control or not.
In particular, this puts under the spotlight an unusual ``quality'' of scattering integrands: \emph{fast numerical evaluation}. While modern developments of amplitudes have mostly been about the discovery of stunningly compact integrands, here the emphasis is not on whether we can write them on the back of an envelope but rather if we can evaluate them at a random point in the blink of an eye.

This paper is structured as follows. 
We warm up with a self-contained review of Feynman and curve integrals in Section \ref{sec:graphandsurfaces}.
We then turn to tropicalization in \ref{sec:tropicalization}, where we introduce the surface Hepp bound.  We also offer a slightly different derivation of the recursive formula presented in \cite{Borinsky:2020rqs} which lends itself to a simple generalization to the case of surfaces.
The main result of the paper is presented in Section \ref{sec:sampling}, where we turn to the problem of sampling. We describe in details the direct Feynman approach and the more efficient dual sampling algorithm. We also summarize the implementation provided as an ancillary file. 
In Section \ref{sec:samplingproblem} we describe the optimization strategy to extend our results beyond the limited setting considered in the paper.
Finally, in Section \ref{sec:conclusions} we draw our conclusions, discuss the next steps and highlight some interesting questions suggested by the ideas put forward in this paper.

\section{Graphs and Surfaces}
\label{sec:graphandsurfaces}

Before we get started, we define precisely the main object of our study: the surface ordered amplitudes in $\mathrm{Tr\ } \phi^3$ theory.

\subsection{Feynman Integrals}

We warm-up with a short review of Feynman integrals in parametric form. 

A graph, $G$, is a collection of finitely many vertices and edges. An (oriented) edge $e = (v_1,v_2)$ is an (ordered) pair of vertices, which may be the same. 
We denote by $-e = (v_2,v_1)$ the opposite orientation. We also say that the vertices $v_1,v_2$ and the edge $e$ are adjacent to each other. Note any each edge is adjacent to two vertices, while a vertex may be adjacent to multiple edges. We say that a vertex is external if it is adjacent to a single edge, which we also call external. 
To each oriented edge, $\pm e$, we attach a momentum $q_e^\mu \in \mathbb{R}^D$, with $q_{-e}^\mu= - q_e^\mu$. F
or every internal vertex $v$ we impose the relation,
\begin{align}
    \sum_{e \mathrm{\ incident\ at\ } v} q_e^\mu = 0,
\end{align}
the sum runs over all edges adjacent to the vertex, all oriented to be incoming in the vertex $v$.

Mathematically, the vertex relations define the first homology group of the graph. Physically, they encode the conservation of momenta. 
In particular they imply that the external momenta must satisfy $\sum_{in} q_e = \sum_{out} q_e$.
Solving the vertex relations, all momenta $q_e^\mu$ can be expressed in terms of $n-1$ of the external legs momenta, plus a collection $\ell_1, \dots, \ell_L$ of $L_G$ independent internal edge momenta. In physics these are called loop momenta, and $L_G$ the loop number of the graph. Note that this coincides with the first Betti number of the graph. We say that a graph is a tree if $L_G = 0$.

There are two important operations on graphs, \emph{deletion} and \emph{contraction}, both of which produce a new graph starting from any subset $\gamma$ of the edges of a graph $G$.
Deletion gives the graph $G\setminus\gamma$, defined to be the graph obtained by removing $\gamma$ from the set of edges of $G$, while leaving the vertex set the same. 
Contraction also removes $\gamma$ from the edges of $G$, but then identifies all vertices adjacent to the same connected component of $\gamma$, yielding the graph $G/\gamma$.

The Feynman integral associated to a connected graph is given by
\begin{align}
    I^{\rm Feyn}_G = \int_{\mathbb{R}^E_{\ge 0}} \bigwedge_{e \in \mathrm{Edges}(G)}d\alpha_e\ \mathcal{U}_G^{-D/2} \exp\left(-\frac{\mathcal{F}_G}{\mathcal{U}_G}\right).
    \label{eq:parametricFeynman}
\end{align}
For a graph with factorized in multiple connected components, say $G = G_L \times G_R$, we define $I_{G_L \times G_R} = I_{G_L}\times I_{G_R}$. The origin of this factorization, is the loop momentum space representation of $I^{\rm Feyn}_G$, from which \eqref{eq:parametricFeynman} is implied by the gaussian integral formula.

\begin{figure}
\centering
\begin{subfigure}{.3\textwidth}
  \centering
  \includegraphics[width=.7\linewidth]{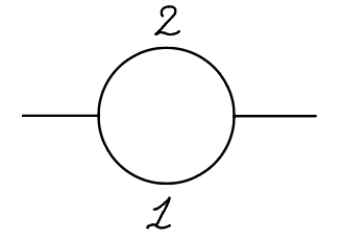}
\end{subfigure}%
\begin{subfigure}{.33\textwidth}
  \centering
  \includegraphics[width=\linewidth]{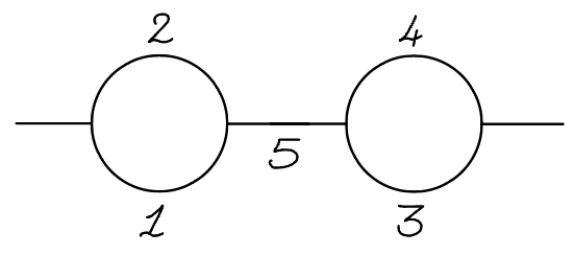}
\end{subfigure}
  \caption{\emph{Examples.} A bubble (Left) and a chain of two bubbles (Right)}
\label{fig:bubbles}
\end{figure}

The key ingredients of \eqref{eq:parametricFeynman} are the first and second Symanzik polynomials, $\mathcal{U}_G$ and $\mathcal{F}_G$.
They are polynomials in the integration variables, of degree $L_G$ and $L_G+1$ respectively. 
They can be defined in two ways, which are equivalent thanks to the matrix-tree theorem.
For the purpose of this paper it will be sufficient to adopt the following definition. 
A $k$-cut of $G$ is a subset $C$ of its edges such that deleting $C$ produces a graph $G\setminus C$ with $k$ connected components, each of which is a tree graph. We denote by $\mathrm{Cuts}_k(G)$ the set of $k$-cuts of $G$.
Note that a $k$-cut partitions the set of external legs into $k$ subsets of legs adjacent to the same connected component of $G\setminus C$.
When $k=2$ we define $s_C$ as the square of the total momentum carried by either subset. It does not matter which subset we choose, thanks to momentum conservation.
Having established this notation, we are ready to define the Symanzik polynomials.
We have
\begin{align}
    \mathcal{U}_G = \sum_{C \in \mathrm{Cuts}_1(G)} \prod_{e \in C} \alpha_e.
    \label{eq:symu}
\end{align}
Define
\begin{align}
    \mathcal{F}^0_G = \sum_{C \in \mathrm{Cuts}_2(G)} s_C\prod_{e \in C} \alpha_e,
    \label{eq:symf0}
\end{align}
then
\begin{align}
    \mathcal{F}_G =\mathcal{F}_G^0 + \left(\sum_{e \in \mathrm{Edges}(G)}m^2_e \alpha_e \right)\mathcal{U}_G.
    \label{eq:symf}
\end{align}
Note that $\mathcal{U}_G$ only depends on the topology of $G$, so the dependence of $I^{\rm Feyn}_G$ on the kinematics enters in \eqref{eq:parametricFeynman} solely via the second Symanzik polynomial.
In \eqref{eq:symf}, we have tacitly adopted Euclidean kinematics by flipping the sign of the mass $m^2$ compared to the standard physical convention.
\begin{example}
    Consider the two graphs shown in Figure \ref{fig:bubbles}.
    If $G$ is the one-loop graph on the left, then we have $\mathcal{U}_G=\alpha_1+\alpha_2$ and $\mathcal{F}_G = (p_1)^2 \alpha_1 \alpha_2 +m^2 (\alpha_1+\alpha_2)\ \mathcal{U}_G$.
    Now let $G$ be the two-loop graph on the right. Then $\mathcal{U}_G = \mathcal{U}_{\gamma_{1,2}}\ \mathcal{U}_{\gamma_{3,4}}$, $$\mathcal{F}^0_G =p_1^2\left(\alpha_1 \alpha_2 \mathcal{U}_{\gamma_{3,4}}+ \alpha_3 \alpha_4 \mathcal{U}_{\gamma_{1,2}} +\alpha_5\ \mathcal{U}_G \right),$$ and
    $$\mathcal{F}_G = \mathcal{F}^0_G+ m^2 \left(\sum_{i=1}^5 \alpha_i\right)\mathcal{U}_G.$$
\end{example}

Symanzik polynomials satisfy a crucial recursive relation, the deletion-contraction property. For every edge $e$ of $G$, we have
\begin{align}
    \mathcal{P}_G = \alpha_e \mathcal{P}_{G\setminus e} + \mathcal{P}_{G/e},
    \label{eq:cutdeletion}
\end{align}
where $\mathcal{P}$ is either $\mathcal{U}$ or $\mathcal{F}$.

Rescaling \emph{all} the edge parameters $\alpha_e$ by a common scale $\alpha_0$, and then integrating over $\alpha_0$, we get to the following equivalent presentation of a Feynman integral as a \emph{projective integral},
\begin{align}
    I^{\rm Feyn}_G = \Gamma(d_G) \int_{\mathbb{P}^{E_G-1}_{\ge 0}} \frac{d\alpha}{\mathrm{GL}(1)}\ \mathcal{U}_G^{d_G-D/2} \mathcal{F}_G^{-d_G}.
    \label{eq:parametricFeynman2}
\end{align}
We have introduced a shorthand notation for the standard volume form,
\begin{align}
    d\alpha =\sum_{e=1}^{E_G} (-1)^e d\alpha_1 \dots d\hat{\alpha}_e \dots d\alpha_{E_G},
\end{align} 
on the projective space $\mathbb{P}^{E_G-1}$ having $\alpha_e$ as homogeneous coordinates.  
The parametric integral \eqref{eq:parametricFeynman2} is now over the non-negative part $\mathbb{P}_{\ge 0}^{E_S}$ of the projective space, defined by $\alpha_e \ge 0$.
In practice the integral can be evaluated by setting one of the $\alpha_e$ to $1$, which is indicated by the notation $\frac{1}{\mathrm{GL}(1)}$.
The quantity $d_G \coloneqq E_G -L_G \frac{D}{2}$ is known as \emph{superficial degree of divergence} of $G$.

With the benefit of hindsight, we find it convenient to strip the prefactor $\Gamma(d_G)$ and define
\begin{align}
    I_G \coloneqq \int_{\mathbb{P}^{E_G-1}_{\ge 0}} \frac{d\alpha}{\mathrm{GL}(1)}\ \mathcal{U}_G^{d_G-D/2} \mathcal{F}_G^{-d_G}.
    \label{eq:parametricFeynman3}
\end{align}
It is important to note that, unlike $I^{\rm Feyn}_G$, $I_G$ does not factorize on disconnected graphs,
\begin{align}
    I_{G_L \times G_R} = \frac{\Gamma(d_{G_L})\Gamma(d_{G_R})}{\Gamma(d_{G_L}+d_{G_R})} I_{G_L} \times I_{G_R}.
    \label{eq:almostfactgraphs}
\end{align}

This concludes our review of Feynman integrals, we next turn to curve integrals, by replacing edges of a graphs with curves on surfaces.

\subsection{Curve Integrals}

We now quickly review the features of curve integrals that are relevant for the present paper. Further details and illustrative examples can be found in \cite{counting1}, and even further will be presented soon in upcoming work.

Curve integrals are a novel representation of scattering amplitudes as a single integral taken over the space of curves that can be drawn on some surface $S$.
This space is also known as the \emph{tropical Teichm\"uller space} \cite{fock2005dualteichmullerlaminationspaces}.
At this stage the surface $S$ itself is nothing but a label for a specific term in a perturbative series expansion for the scattering amplitude.
This is the famous t'Hooft  large $N$ expansion \cite{1973H}, an amplitude involving $n$ external particles can be written as
\begin{align}
    A_n = \sum_{S \in \mathrm{Surfaces}_n} C_S \times A_S,
    \label{eq:surfaceordering}
\end{align}
where the sum runs over all surfaces with $n$ marks on their boundary. $C_S$ is a quantity built out of the coupling constants, the order of the gauge group, as well as by traces of products of generators of the group.
Its precise value can be derived straightforwardly by Feynman rules, but it is not important here.
In \eqref{eq:surfaceordering}, the dependence of the amplitude $A_n$ on the external kinematics is completely disentangled from the color dependence. 
The kinematics enters only through the factors $A_S$, the \emph{surface-ordered} amplitudes.
These are the quantities that we set out to evaluate numerically.

\begin{figure}[h!]
    \centering
    \includegraphics[width=0.5\linewidth]{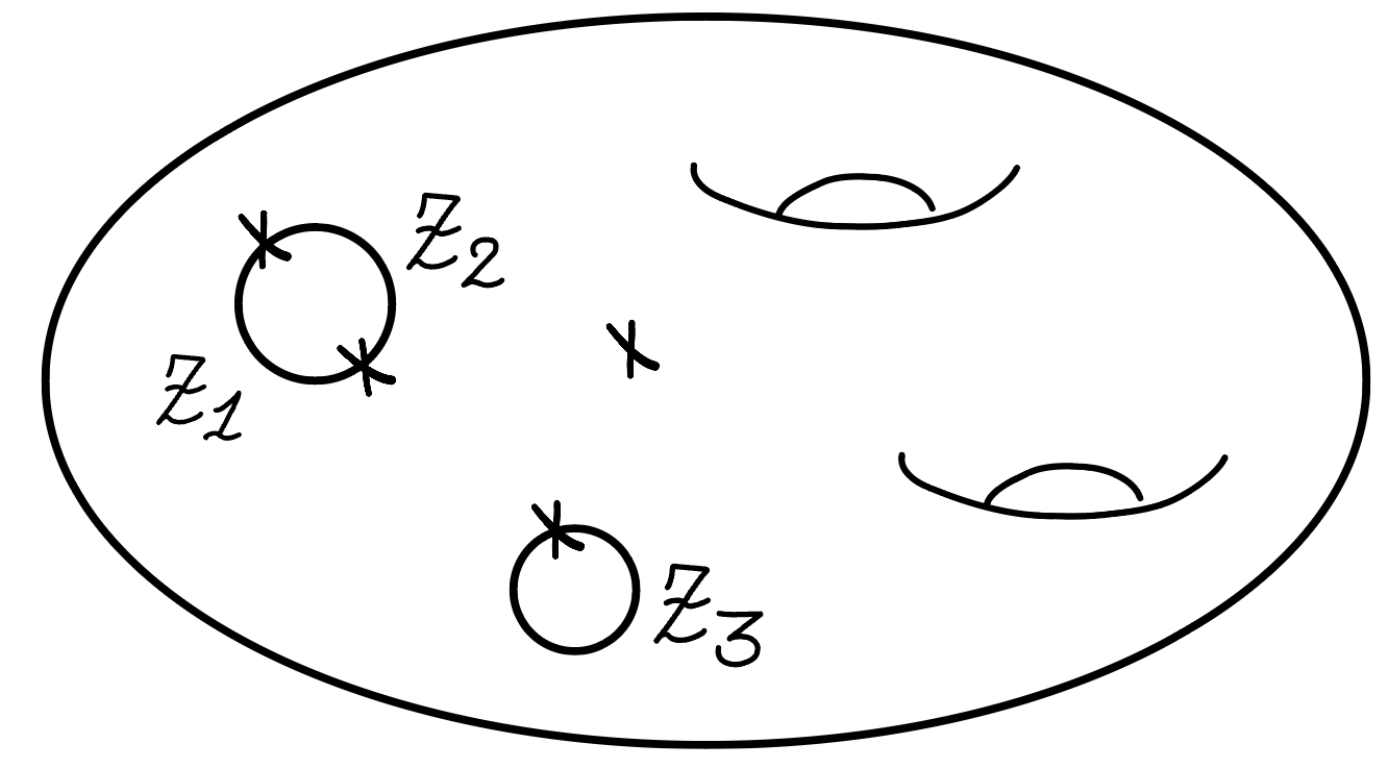}
    \caption{\emph{Surface labelling.} The surface $S = \{2,\{[z_1,z_2],[z_3],[\emptyset ]\}\}$. The corresponding unlabelled version is $\tilde{S} = \{2,\{2,1,0\}\}$}.
    \label{fig:surfacelabel}
\end{figure}

What is a surface? A connected, orientable, bordered, surface is uniquely identified by a small amount of data describing its topological invariants. These are the genus and a partition of its boundary into consecutive boundary segments, separated by marks.
A boundary component which is not broken in segments and marks is represented as a single puncture in the bulk of the surface.
It is useful to also define \emph{unlabeled surfaces}.
These are obtained from an ordinary surface by forgetting the labels of the boundary segments, and instead recording only how many are chained up to form each of the boundary components of $S$.
At the risk of being over pedantic, let us describe precisely how the labeling is defined.
If we denote sets as $\{ \dots \}$ and cyclically ordered sets as $[ \dots ]$, then a labeled surface is $S = \{g,\{[m_{1,1},\dots], \dots,[m_{b,1, \dots}]\}\}$, see Figure \ref{fig:surfacelabel} for an example.

\begin{figure}[h]
    \centering
    \includegraphics[width=0.15\linewidth, angle=-90]{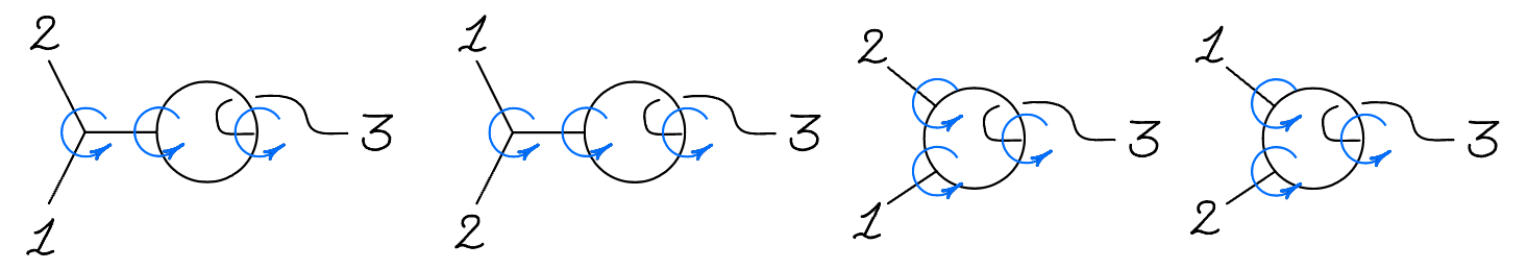}
    \caption{\emph{Fatgraphs.} The fatgraphs contributing to $A_{\{0,\{2,1\}\}}$. The orientation of the edges adjacent at each vertex is shown in blue.}
    \label{fig:fatgraphsA21}
\end{figure}

The surface-ordered amplitude $A_S$ is defined by summing over all the graphs that can be embedded in $S$ with the external legs anchored at the boundary segments and without self-intersections. 
These can also be described as the graphs that are dual to a triangulation of the surface.
Note that the orientation of the surface automatically induces an orientation of the edges adjacent to any vertex of the graph. 
A graph enriched by this combinatorial data is called a \emph{fatgraph}, and accordingly we denote by $\mathrm{FatGraphs}(S)$ the set of all fatgraphs associated to a surface $S$.
\begin{example}
    Let $S=\{0,\{2,1\}\}$ be an annulus with two marks on one boundary component and a single mark on the other.
    In Figure \ref{fig:fatgraphsA21} are shown all fatgraphs in $\mathrm{FatGraphs}(S)$. Note that distinct fatgraphs may descend to the same graph when the orientation of the edges at a vertex is forgotten, and in particular contribute the same value to $A_S$. 
\end{example}

Fatgraphs appear naturally in gauge theories, due to the matrix nature of the fields in color space.
Each fatgraph contributes to the amplitude via the standard Feynman rules.
In the case of scalar theories, such as $\mathrm{Tr}(\phi^3)$ theory, the result can be directly written as a sum of parametric Feynman integrals associated to the underlying graphs,
\begin{align}
    A_S = \sum_{\Gamma \in \mathrm{FatGraphs}(S)} I^{\rm Feyn}_\Gamma =  \sum_{\Gamma \in \mathrm{FatGraphs}(S)} \Gamma(d_\Gamma) \int_{\mathbb{P}^{E_\Gamma-1}_{\ge 0}} d\alpha_\Gamma\ \mathcal{U}_\Gamma^{d_\Gamma-\frac{D}{2}} \mathcal{F}_\Gamma^{-d_\Gamma}.
    \label{eq:curveint0}
\end{align}
The surprise is that there is a global parametric formula for $A_S$, the \emph{curve integral}
\begin{align}
    A_S = \Gamma(d_S)\int_{\mathbb{P}^{E_S-1}} \frac{\omega}{\mathrm{GL}(1) \times \mathrm{MCG}_S}\ \mathcal{U}_S(\alpha_C)^{d_S-\frac{D}{2}}\ \mathcal{F}_S(\alpha_C)^{-d_S}.
    \label{eq:curveint}
\end{align}
Crucially, this is \emph{not} merely obtained by exchanging the sum and integral signs in \eqref{eq:curveint0}. 
There is an additional structure that tells us how the Feynman simplices miracolously glue together into the curve integral.
We will illustrate piece by piece the ingredients that go in the definition of a curve integral.

\begin{figure}[h!]
    \centering
    \includegraphics[width=0.75\linewidth]{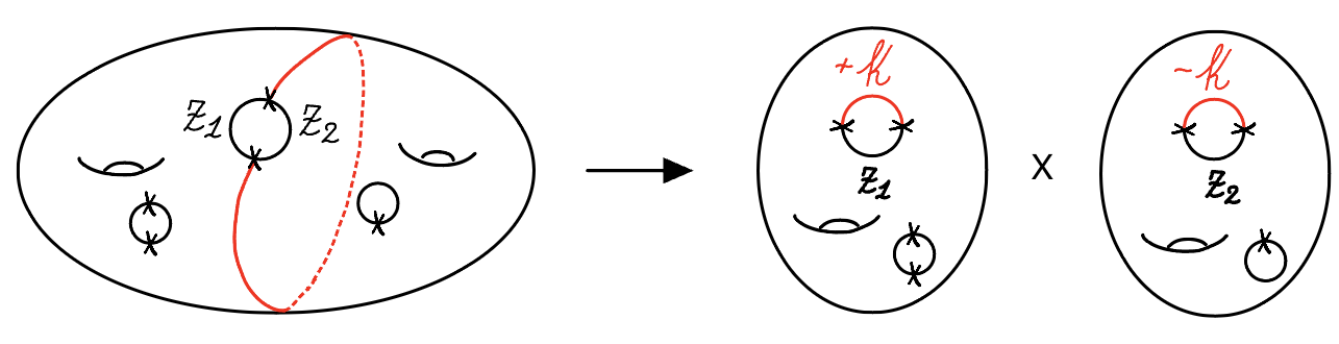}
    \caption{\emph{Tree curves. } A curve, when cut, may factorize the surface in two. The loop order is unchanged.}
    \label{fig:factorization}
\end{figure}

\begin{figure}[h!]
    \centering
    \includegraphics[width=0.12\linewidth,angle=270]{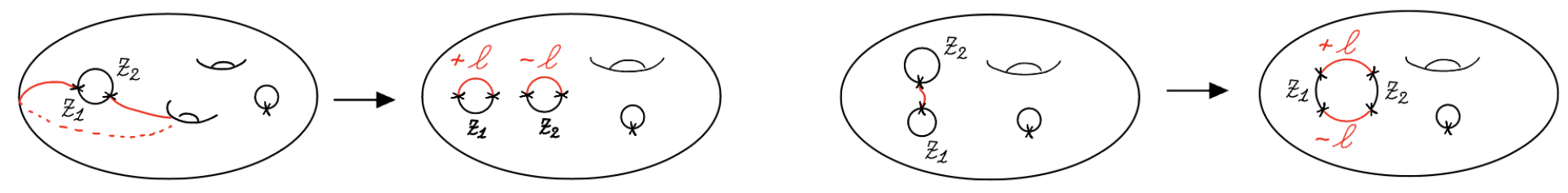}
    \caption{\emph{Loop curves. }A curve, when cut, may decrease the genus (Left) or merge two boundaries (Right). The loop order decreases.}
\label{fig:loop}
\end{figure}

The stars of the show are the curves that we can draw on a surface $S$, up to homotopy, with endpoints at the marks of the surface, and with no self-intersections.  
The boundary segments themselves are a special class of curves: because they are dual to external legs they do not contribute the the curve integrand.
The set of remaining curves will be denoted by  $\mathrm{Curves}(S)$.
Curves are naturally partitioned in two sets.
We say that a curve is \emph{factorizing} if cutting along it produces a disconnected surface (see Figure \ref{fig:factorization}), otherwise we call it non factorizing (see Figure \ref{fig:loop}).

Any curve defines an element of the homology of the surface relative to the marks. As in the case of graphs, we can find a basis for the homology, and expand every curve on it.
The basis consists of the special external curves, i.e. the external boundary segments, as well as an arbitrary choice of $L_S$ non-factorizing curves.
Here, $L_S$ is the same as the loop number of \emph{any} fatgraph that can be embedded in $S$.
In physics, the homology plays the role of the momentum carried by the curve.
An alternative way to describe the partition of curves into factorizing and non-factorizing, is that the factorizing curves are those that carry a subset of the external momenta and no loop momenta.


To each curve $C$ there is an associated vector $g_C \in \mathbb{R}^{E_S}$ that uniquely identifies the curve \cite{counting1,fock2005dualteichmullerlaminationspaces}. Here, $E_S$ is the maximal number of curves that can be drawn simultaneously on $S$ without intersections\footnote{If $S=\{g,\{n_1,\dots,n_b\}\}$, then $E_S=6(g-1)+3 b + \sum_{i=1}^b n_i$ and $L_S = 2 g+b-1$.}.
We construct a \emph{polyhedral fan} $\Sigma_S$ by taking the positive spans of maximal collections of vectors $\{g_{C_1}, \dots , g_{C_m}\}$ corresponding to curves that can be drawn simultaneously without intersections on $S$. This data specifies the homotopy class of a triangulation of $S$.
Dually, it also defines a Feynman fatgraph $\Gamma$ embedded on $S$. For this reason the fan $\Sigma_S$ was called \emph{Feynman fan} in \cite{counting1}.

To every curve we also associate a \emph{headlight function}, $\alpha_C(t)$. It is defined by imposing $\alpha_C(g_{C'}) = \delta_{C,C'}$, where $\delta$ is the Kronecker symbol, and then extended piecewise linearly to the entire Feynman fan. The ``second miracle'' of surfaceology is that there is a much more efficient way to calculate the headlight function $\alpha_C$ by a matrix formula. Somewhat disappointingly, however, this miracle does not play any role in the present paper. 
We interpret it optimistically as the clue that a much better sampling algorithm could exist, a point on which we will touch again in the conclusions.

The \emph{canonical form} $\omega$ is particularly simple when written using the same parametrization as for the headlight functions $\alpha_C(t)$: $\omega = dt_1\dots dt_{E_S}$. 
It can also be described as a piecewise differential form with respect to the headlight variables $\alpha_C$: to a cone $\sigma$ formed by the positive span of vectors $\{g_{C_1}, \dots,g_{C_{E_S}}\}$, and labeled by a fatgraph $\Gamma$, we have $\omega|_{\sigma} = d\alpha_1 \dots d\alpha_{E_S}$\footnote{This follows from the uni-modularity of the Feynman fan: $\det \left( g_{C_1}, \dots,g_{C_{E_S}}\right)=\pm 1$.}. 

The \emph{surface} Symanzik polynomials $\symU_S$ and $\symF_S$ have a definition similar to the standard graph polynomials. 
We define a $k$-cut of the surface $S$, to be any collection of curves, $\mathcal{C} = \{C_1, \dots, C_m\}$, such that cutting $S$ along all the curves $C_i$ yields $k$ disks with marked points, that is to say surfaces of the form $S_j = \{0,\{n_j\}\}$, we denote by $\mathrm{Cuts}_k(S)$ the set of all $k$-cuts of $S$. The $n$ boundary segments of $S$, are partitioned among the boundaries of the disks $S_j$.  If $k=2$, we can define the $s_{\mathcal{C}}$ as the norm of the total momentum carried by the external boundary segments on either disk.

\begin{figure}
    \centering
    \includegraphics[width=0.75\linewidth]{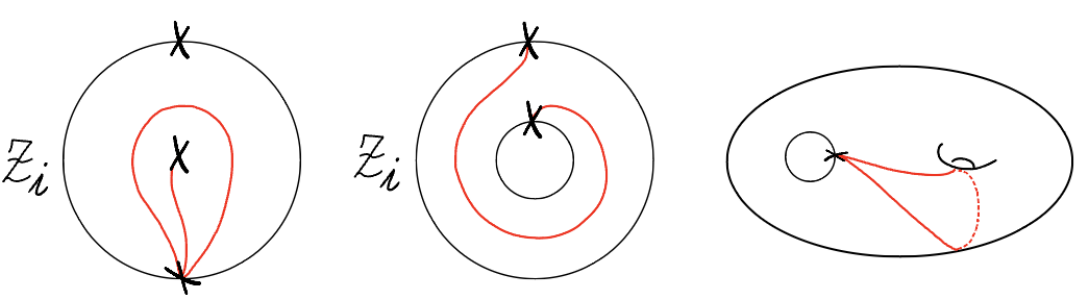}
    \caption{\emph{Curves on surfaces. } Curves on $S=\{0,\{2,0\}\}$ (Left), on $S=\{0,\{1,1\}\}$ (Middle), on $S=\{1,\{1\}\}$(Right).}
    \label{fig:planarannulus}
\end{figure}

Then we set,
\begin{align}
    \symU_S \coloneqq \sum_{\mathcal{C} \in \mathrm{Cuts}_1(S)}\prod_{C \in \mathcal{C}} \alpha_C,
\end{align}
\begin{align}
    \symF^0_S \coloneqq \sum_{\mathcal{C} \in \mathrm{Cuts}_2(S)}s_{\mathcal{C}}\prod_{C \in \mathcal{C}} \alpha_C,
\end{align}
and
\begin{align}
    \symF_S \coloneqq \mathcal{F}^0_S + \left(\sum_{C \in \mathrm{Curves}(S)} m^2_C\ \alpha_C   \right) \mathcal{U}_S.
\end{align}
Note that these are polynomials if we think of the $\alpha_C$ as \emph{variables}. But in truth the $\alpha_C$ themselves are piecewise-linear functions on the Feynman fan. 
Another subtlety, is that the polynomials have infinitely many terms. Concretely this is not an issue, because we will see shortly that we only ever have to consider the restriction of the polynomials in a domain of the Feynman fan where only finitely many terms are non-vanishing.
\begin{example}
    \label{ex:treelevelcurveint}
    Let $S=\{0,\{n\}\}$ be a disk with $n$ marks on the boundary. 
    Every curve cuts the disk into two connected surfaces, so $$\mathrm{Cuts}_2(S) = \{ \{C\}, C \in \mathrm{Curves}(S)\}.$$
    Due to the fact that we cannot cut a disk into a \emph{single} connected surface, we have $$\mathcal{U}_S=1,$$
    and
    $$\mathcal{F_S}=\sum_{C \in \mathrm{Curves}(S)} \alpha_C s_C.$$
    Therefore, the curve integral for the amplitude is
    $$A_S=\Gamma(n-3)\int_{\mathbb{P}^{n-4}} \frac{\omega \ }{\mathrm{GL}(1)} \mathcal{I}_S = \Gamma(n-3)\int_{\mathbb{P}^{n-4}} \frac{\omega\ }{\mathrm{GL}(1)} \left( \sum_{C \in \mathrm{Curves}(S)} \alpha_C (m^2+ s_C ) \right)^{-(n-3)}. $$
    It can be calculated explictly,
    $$A_S = \sum_{G \in \mathrm{FatGraphs}(S)}I^\mathrm{Feyn}_G = \sum_{\mathcal{C}=(C_1,\dots,C_{n-3}) \in \mathrm{Cuts}_{n-3}(S)} \frac{1}{(m^2+s_{C_1}) \dots (m^2+s_{C_{n-3}})},$$
    which is the usual representation of a tree level scattering amplitude in $\mathrm{Tr}(\phi^3)$ theory.
\end{example}
\begin{example}
    Consider $S = \{0,\{2,0\}\}$, an annulus with two marked points on one of its boundaries.
    There are four possible curves on $S$, some of which are depicted in Figure \ref{fig:planarannulus} (Left). Let $\alpha_1,\alpha_2$ be the variables corresponding to non-factorizing ones and $\alpha_{T_1},\alpha_{T_2}$ the factorizing ones. 
    We have $$\mathrm{Cuts}_2(S) = \{\{\alpha_1, \alpha_{2}\},\{\alpha_1, \alpha_{T_1}\},\{\alpha_2, \alpha_{T_2}\}\}.$$
    Accordingly, $\mathcal{U}_S = \alpha_1 + \alpha_2,$ $$\mathcal{F}_S^0 = s\left( \alpha_1  \alpha_2 + \alpha_{T_1} \mathcal{U}_S+ \alpha_{T_2} \mathcal{U}_S\right),$$ and $$\mathcal{F}_S = \mathcal{F}_S^0 + \left(\sum_{C\in\mathrm{Curves}(S)} m^2 \alpha_C \right)\mathcal{U}_S.$$
\end{example}

There is a group acting on $\mathrm{Curves}(S)$. It is called the mapping class group (MCG) of $S$, and it will be denoted by $\mathrm{MCG}_S$.
Although elementary, the precise definition of $\mathrm{MCG}$ goes beyond the scope of this paper. It suffices here to know that two curves are equivalent up to $\mathrm{MCG}_S$ if and only if the surfaces obtained cutting along them are topologically equivalent.
It follows that we can characterize the equivalence class of a curve $C$, up to $\mathrm{MCG}_S$, by providing the topological data for the surface $S \setminus C$. 
Note that the $S \setminus C$ may be disconnected, in which case we simply label it by giving the topological data for each of its connected components. There are three possible cases to be aware of, depicted in Figures \ref{fig:factorization} and \ref{fig:loop}.

\begin{figure}
    \centering
    \includegraphics[width=0.5\linewidth]{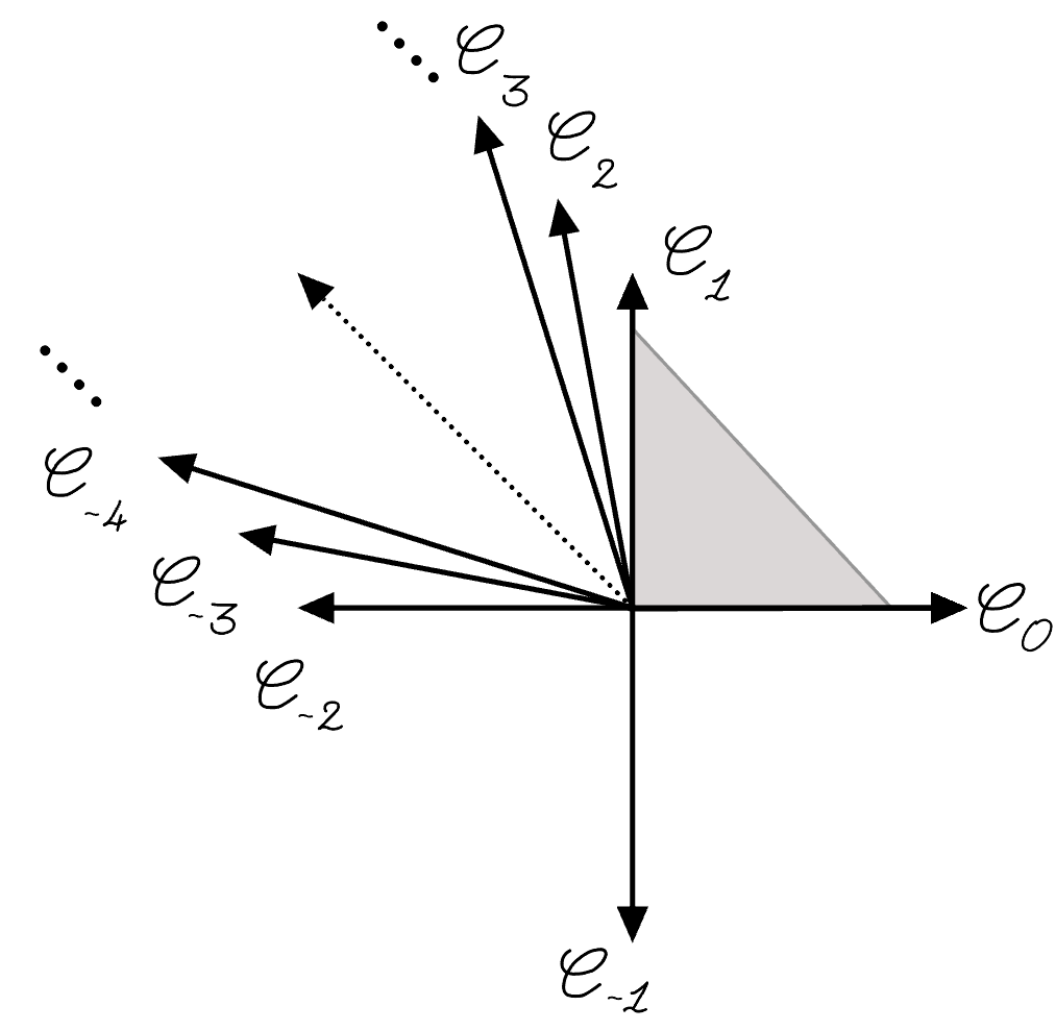}
    \caption{\emph{The Feynman fundamental domain. }The fan of the surface $S=\{0,\{1,1\}\}$ with a Feynman foundamental domain for the MCG highlighted in gray.}
    \label{fig:a11fundamental}
\end{figure}

\begin{example}
    Let $S=\{0,\{1,1\}\}$ be the annulus with one marked point on each boundary shown in Figure \ref{fig:planarannulus} (Middle). There are infinitely many curves on $S$, but they all cut $S$ down to a disk with four marked points on the boundary, and therefore they are all equivalent up to MCG.
    Let $S^{\rm torus}=\{1,\{1\}\}$,  be the torus with one marked point on its boundary show in Figure \ref{fig:planarannulus} (Right).
    There is only one curve up to MCG on it, which cuts the torus to the annulus $S = \{0,\{2,1\}\}$.
\end{example}
The action of $\mathrm{MCG}_S$ is purely combinatorial on $\mathrm{Curves}(S)$, but it is translated to a piecewise linear action on $\mathbb{R}^{E_S}$.
The action is defined by $\gamma(g_C) \coloneqq g_{\gamma(C)}$, for $\gamma \in \mathrm{MCG}_S$, and is then extended piecewise linearly with respect to the Feynman fan $\Sigma_S$.
The notation $\frac{1}{\mathrm{MCG}_S}$ means that the integral \eqref{eq:curveint} is restricted to a fundamental domain of this action. 
Concretely, this can be described by listing all fatgraphs $\Gamma \in \mathrm{FatGraphs}(S)$ compatible with the topology of $S$, and then considering any of the cones $\sigma_\Gamma$ dual to each $\Gamma$\footnote{We warn the reader that even if two distinct fatgraphs descend to the same graph, thus contributing the same value to $A_S$, we still need \emph{both} to carve out a fundamental domain of the MCG}. 
The integrand of \eqref{eq:curveint} is MCG-invariant, so that the value of integral does not depend on this choice.
Furthermore, in any such fundamental domain, only finitely many curves $C$ have non-vanishing headlight functions $\alpha_C$. Therefore, the Symanzik polynomials are well defined polynomials in these finitely many headlight functions $\alpha_C$.
\begin{example}
Consider the surface $S=\{0,\{1,1\}\}$, an annulus with one marked point on each boundary component.
The Feynman fan for $S$ is show in Figure \eqref{fig:a11fundamental}.
There is only one diagram that can be drawn on this surface, a one-loop fatgraph that is dual to all the cones of the fan.
Pick any of these cones of $\Sigma_S$, say the one formed by the curves $C_0$ and $C_1$.
In this cone, we have $\mathcal{U}_S = \alpha_{C_0} + \alpha_{C_1}$ and $\mathcal{F}_S = s\  \alpha_{C_0} \alpha_{C_1} + m^2 (\alpha_{C_0}+\alpha_{C_1})^2$, where $s$ is the momentum squared carried by either boundary segment.
\end{example}
Finally, we define the quantity $d_s \coloneqq  E_S - L_S \frac{D}{2}$, where $L_S$ is the loop number of any of the fatgraphs $\Gamma \in \mathrm{FatGraphs(S)}$. It could perhaps be called the \emph{surf}-icial degree of divergence of $A_S$.
The overall prefactor $\Gamma(d_S)$ arise from the same step that brought us from \eqref{eq:parametricFeynman} to \eqref{eq:parametricFeynman2}, that is by rescaling \emph{all} headlight functions by a common scale and integrating it away.
This is why in \eqref{eq:parametricFeynman2} we chose to rescale all Feynman parameters by a common scale, even if they belong to different connected components of the graph or if they do not carry loop momenta \footnote{Such propagators can of course be integrated exactly so it may seem counterintuitive to keep them un-integrated. The reason behind our choice is that it results in the conceptually easiest dual sampling algorithm.}.
Concretely we can gauge-fix the rescaling by localizing the integral over the piecewise locus
\begin{align}
    \mathcal{V}_d \coloneqq \left\{ \sum_{C \in \mathrm{Curves}(S)} \alpha_C =1\right\}.
\end{align}
In this language, the second miracle of surfaceology can be thought of as providing a piecewise linear parametrization for this locus, by which $\alpha_C$ turns into the headlight function $\alpha_C(t)$ on $\mathbb{R}^n$. 

In general, curve integrals may be divergent and thus need to be regulated. This can be done by familiar dimensional regularization. That is we treat $D$, which physically is the space-time dimension, as a parameter on which $A_S$ depends. The divergences of $A_S$ manifest as poles around integer values for $D$.

In the following we will consider a situation where $A_S$ is directly finite. 
To do so, we will restrict ourselves to consider surfaces without punctures, in $D=2$ space-time dimensions, and assigning a mass to all curves.  The mass protects from infrared divergences. 
An elementary calculation shows that in $D=2$ the superficial degree of divergence of a graph of  $\mathrm{Tr}(\phi^3)$ theory with $v_i$ internal vertices is $v_i-1$, thus only one-loop tadpoles have (logarithmic) ultraviolet divergences. Such diagrams can only be drawn on a punctured surface, see Figure \ref{fig:tadpole}.

Restricting to finite amplitudes only represents a severe limitation to the ideas presented in this paper, but we will argue in the conclusions why we are optimistic that it will not constitute an insurmountable one.

\section{Tropicalization}
\label{sec:tropicalization}

In this section we introduce a central character of our construction, the Hepp bound.
We begin by reviewing the case of Feynman integrals, associated to individual graphs, and then generalize to curve integrals.

\subsection{Single Graphs}

The Hepp bound for a Feynman integral was defined in \cite{Panzer:2019yxl}. In \cite{Borinsky:2020rqs}, a recursive formula satisfied by the Hepp bound was brilliantly used to numerically evaluate Feynman integrals by a tropical Monte Carlo algorithm.
In a nutshell, the Hepp bound is defined by replacing the Symanzik polynomials by their \emph{tropicalization}. 
Let us introduce the monomial notation $\alpha^{\bf m} \coloneqq \prod_{e=1}^E \alpha_e^{m_e}$, so that an arbitrary polynomial can be written as
$P = \sum_{{\bf m} \in \mathcal{P} \subset \mathbb{Z}^n} s_{\bf m} \alpha^{\bf m}$. Then we define $P^{\rm tr}\coloneqq \max_{m \in \mathcal{P}} \alpha^{\bf m}$. At first this may look like a strange operation, but in fact tropicalization has a very physical meaning, and it is so ubiquitous in physics because it shows up when taking limits in logarithmic coordinates. 
For any function $f : \mathbb{R}^E_{\ge 0} \to \mathbb{R}$, let us define a new function $\mathrm{Trop\ f} : \mathbb{R}^E \to \mathbb{R}$ via
\begin{align}
    [\mathrm{Trop\ }f] (t) \coloneqq \lim_{\alpha' \to 0^+} \alpha' \log |f(e^{\frac{t}{\alpha'}})|.
\end{align}
Then $f^{\mathrm{tr}}(e^t) = e^{[\mathrm{Trop} f](t)}$.
Note that this definition is applicable well beyond the case of polynomials, including transcendental functions.

The Hepp bound of a graph is defined by
\begin{align}
    H_G = \int_{\mathbb{P}^{E_G-1}_{\ge 0}} \frac{d\alpha}{\mathrm{GL}(1)}\ (\mathcal{U}_G^{\mathrm{tr}})^{d_g-D/2} (\mathcal{F}_G^{\mathrm{tr}})^{-d_G}.
    \label{eq:heppgraph}
\end{align}
It is not difficult to see that, as long as the coefficients $\mathcal{F}_G$ are positive, the parametric integrand appearing in the definition of $I_G$ is bounded from below and above by the Hepp integrand. In this sense, it is not surprising that the Hepp bound $H_G$ should provide an estimate for the value of $I_G$. What is striking, is how accurate this estimate turns out to be \cite{Panzer:2019yxl}.

In simple cases, the Hepp bound can be evaluated immediately.
\begin{example}
Let $G$ be an arbitrary tree graph, with $n$ edges, then
\begin{align}
    H_G = \int_{\mathbb{R}^{n-1}_{\ge 0}} \frac{d\alpha_1 \dots d\alpha_{n-1}}{\max(\alpha_1,\dots,\alpha_{n-1},1)^n} = n.
\end{align}
\end{example}

In order to calculate more complicated Hepp bounds, the first step is to understand the tropicalization of the Symanzik polynomials.
This in turn requires to introduce the notion of \emph{Newton polytope}.
We refer to \cite{Salvatori:2024nva} for an in-depth review of these ideas and for precise definitions. Here we content ourselves with reporting a few essential facts.
For any polynomial $P$, its tropicalization is a function that is piecewise linear on the normal fan of the Newton polytope $\mathcal{P} = \mathrm{Newt} P$.
Furthermore, if a vector $\rho$ appears in one of the inequalities carving out $\mathcal{P}$, as in
\begin{align}
    \mathcal{P} \subseteq \{\alpha \in \mathbb{R}^{n}| d_\rho - \rho \cdot \alpha \ge 0\},
\end{align}
where $d_\rho \in \mathbb{R}$ is a constant, then $[\mathrm{Trop}\ P](\rho) = d_\rho$.
From these facts, it follows immediately that if $\mathrm{Newt\ P} \subseteq \mathrm{Newt\ }P'$, we have that $\mathrm{Trop}(P+P') = \mathrm{Trop}(P')$.

Having spelled out these general facts about tropicalization, we focus on the case of Symanzik polynomials.
The polytope $\mathrm{Newt\ } \mathcal{U}_G$ is completely understood. It is a generalized permutohedron associated to the supermodular function $z_\gamma = L_\gamma$ \cite{Schultka:2018nrs}.
In other words, it is a polytope with facet description
\begin{align}
\mathrm{Newt\ } \mathcal{U}_G = \{\alpha \in \mathbb{R}^{E_G}|-L_\gamma - \rho_\gamma \cdot \alpha \ge 0,  \ \forall\ \gamma \subseteq G\},
\label{eq:newtu}
\end{align}
where $\gamma$ denotes an arbitrary edge subgraph of $G$ and the inequality is always saturated for $\gamma = G$.
The vector $\rho_\gamma \in \mathbb{R}^E$ is an indicator for the subgraph $\gamma$: its entries are labeled by the edges of $G$ and take the value $\rho_{\gamma,e} = - 1$ if $e \in \gamma$ and $0$ else.
Furthermore, the normal fan of $\mathrm{Newt\ } \mathcal{U}_G$ admits as a simplicial refinement the normal fan of the permutohedron built on the set of edges of $G$, $\mathcal{P}_\mathrm{Edges (G)}$.
It means that the normal fan of $\mathrm{Newt\ } \mathcal{U}_G$ is triangulated by simplices --- or \emph{sectors} --- $\Delta_\sigma$ that are in bijection with all the possible orderings of the edge variables $\alpha_e$, i.e. with all permutations of the edges of $G$.
The (generators of the) rays of the cone $\Delta_\sigma$ are described in terms of the permutation $\sigma$ as follows.
Let $e_1, \dots, e_{E_G}$ be the edges of the graph $G$ and 
consider the subgraphs $\gamma_{\sigma,i} \coloneqq \{e_{\sigma_1}, \dots, e_{\sigma_i}\}$ with $i=1,\dots,E_G-1$.
Then $\mathrm{Rays\ } \Delta_\sigma = \{\rho_{\gamma_1}, \dots,\rho_{\gamma_{E_G-1}}\}$, where $\rho_\gamma$ is as in \eqref{eq:newtu}.

The second Symanzik polytope is much less understood. Some essential questions regarding its facet description remain unanswered, if not in special cases \cite{Schultka:2018nrs}.
In our study, we will consider only the case where all propagators are massive, $m^2_e \ne 0$, in which case an important simplification takes place. From the formulae \eqref{eq:symu},\eqref{eq:symf0} and \eqref{eq:symf}, it is clear that
\begin{align}
   \mathrm{Newt\ }\mathcal{F}_G \subseteq \mathrm{Newt\ }\left(\sum_{e \in \mathrm{Edges}(G)} \alpha_e\right)\mathcal{U}_G, 
\end{align}
so that 
\begin{align}
   \mathrm{Trop\ }\mathcal{F}_G &= \mathrm{Trop\ }\left(\sum_{e \in \mathrm{Edges}(G)} \alpha_e\right)\mathcal{U}_G= \left(\max_{e \in \mathrm{Edges}(G)} \alpha_e\right) + \mathrm{Trop\ }\mathcal{U}_G.
   \label{eq:tropfsimplified}
\end{align}

These basic observations already bring us to a simpler formula for the Hepp bound,
\begin{align}
    H_G = \int_{\mathbb{P}^{E_G-1}_{\ge 0}} \frac{d\alpha}{\mathrm{GL}(1)}\ (\mathcal{U}_G^{\mathrm{tr}})^{-D/2} \left(\max_{e \in \mathrm{Edges}(G)}\alpha_e\right)^{-d_G}.
    \label{eq:Hepp2}
\end{align}
In particular, note that the Hepp bound is independent on the kinematical configuration of the graph, i.e. on whether $s_C = 0$  for some 2-cut $C$.
Furthermore, it turns out that the integral in \eqref{eq:Hepp2} can be performed exactly, resulting in a purely combinatorial formula \cite{Panzer:2019yxl}
\begin{align}
    H_G = \sum_{\sigma}  \prod_{\rho_\gamma \in \mathrm{Rays\ } \Delta_\sigma} \frac{1}{d_{\gamma}}
    \label{eq:HeppSec},
\end{align}
where the sum runs over all orderings of the edges.
This can be proven by \emph{geometric sector decomposition} \cite{Kaneko_2010}, that is by decomposing the domain of integration into the images of the simplices $\Delta_\sigma$ under the coordinate-wise exponent map. A detailed discussion can be found for example in \cite{Schultka:2018nrs}.
When restricted to any sector, the integrand reduces to a monomial and can be integrated explicitly resulting in the RHS of \eqref{eq:HeppSec}. 
Let us vigorously stress that in the proof of \eqref{eq:HeppSec} a critical role is played by \eqref{eq:tropfsimplified} and, more importantly, by the knowledge of the facet description \eqref{eq:newtu}.
\begin{figure}
    \centering
    \includegraphics[width=0.5\linewidth]{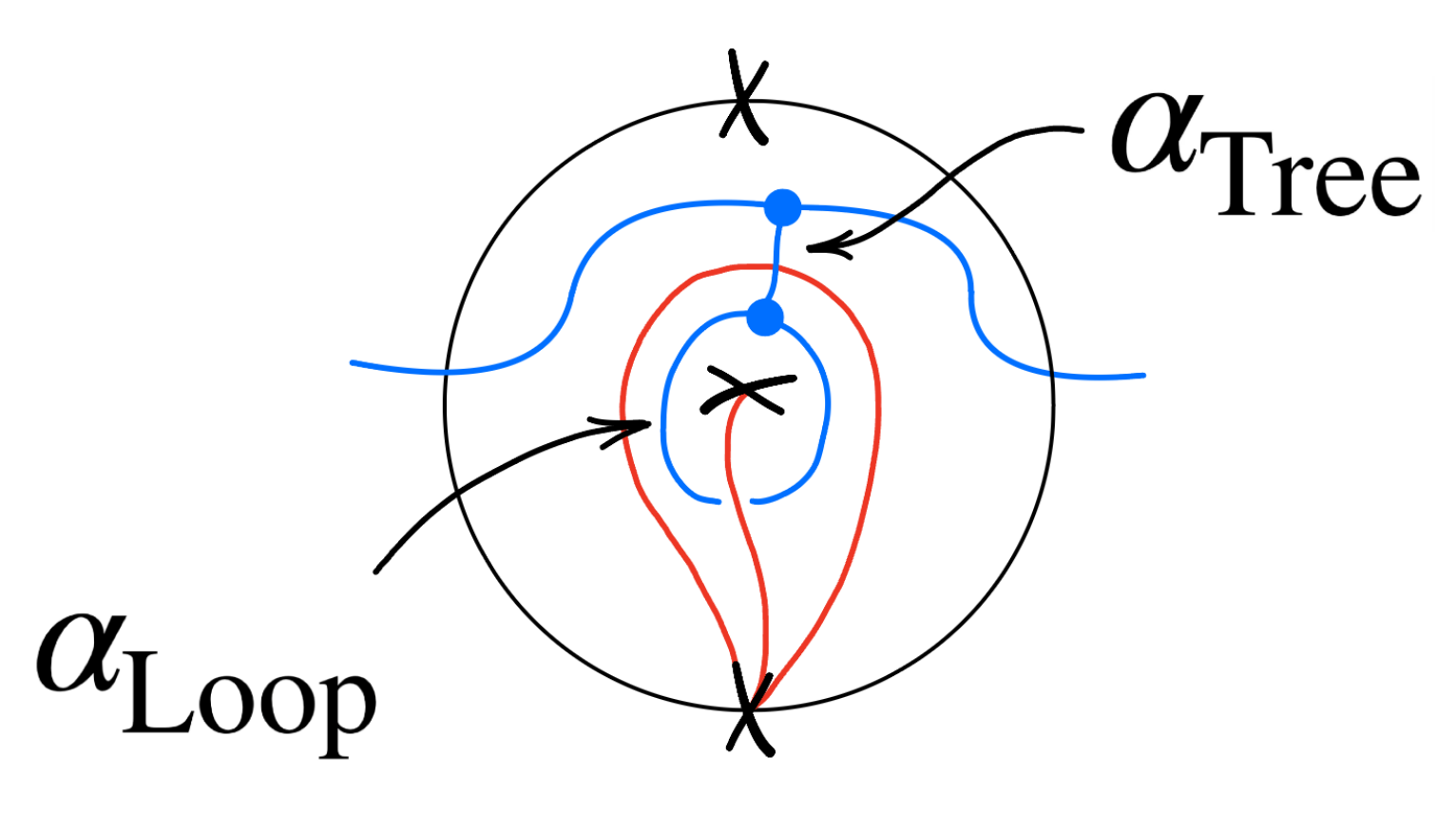}
    \caption{\emph{Tadpoles are punctures.} A ``tadpole'' fatgraph on a punctured surface. It is dual to a triangulation formed by two curves, one is factorizing, the other carries a loop.}
    \label{fig:tadpole}
\end{figure}
\begin{example}
    Consider the ``tadpole'' fatgraph $\Gamma$ that is dual to the triangulation shown in Figure \ref{fig:tadpole}. Let $\alpha_{\rm tree}$ and $\alpha_{\rm loop}$ be the headlight variables associated to the factorizing and non-factorizing curve, respectively.
    We can compute $H_\Gamma$ by fixing $\alpha_{\rm loop}=1$ and by breaking the integration over $\alpha_{\rm tree}$ into sectors $1 > \alpha_{\rm tree} >0$ and $\alpha_{\rm tree} > 1$. 
    In $D=2-2\epsilon$ space-time dimensions we have,
    \begin{align}
        H_\Gamma &= \int_{0}^\infty d\alpha_{\rm tree} \max(1,\alpha_{\rm tree})^{-(1+\epsilon)}= \frac{1}{d_{\gamma_{\rm tree}}} + \frac{1}{d_{\gamma_{\rm loop}}}  = 1 + \frac{1}{\epsilon},
    \end{align}
    where $\gamma_{\rm tree}$ and $\gamma_{\rm loop}$ are one-edge subgraphs, each formed by exactly one of the red curves.
\end{example}

An unsatisfactory feature of the explicit formula is that it involves a sum over factorially many terms, which makes it unpractical to use.
Luckily, there is a more efficient formula to compute the Hepp bound, which was found in \cite{Borinsky:2020rqs}.
We have
\begin{align}
    H_G = \sum_{e \in \mathrm{Edges}(G)} \frac{H_{G\setminus e}}{d_{G\setminus e}}.
    \label{eq:borinsky}
\end{align}
Which justifies introducing the quantity $\hat{H}_G \coloneqq \frac{H_G}{d_G}$. Note that the formula is recursive with respect to the number of edges. 
\begin{example}
    Let $\Gamma$ be the tadpole fatgraph of the previous example. We have $H_{\Gamma} = \hat{H}_{\Gamma\setminus e_{\rm loop}}+\hat{H}_{\Gamma\setminus e_{\rm tree}}= 1 +\frac{1}{\epsilon}$.
\end{example}
\begin{example}
    Let $G$ be a tree with two edges. Then 
    $H_G = \hat{H}_{e_1} + \hat{H}_{e_2} = 2$.
    So $\hat{H}_G = 1$. 
    By induction one easily prove that for an arbitrary tree graph $\hat{H}_G = 1$.
    Next, let $G$ be an $n$-gon, i.e. a one loop graph with $n$ edges. Then
    $H_G = \sum_{e \in \mathrm{Edges}(G)}\hat{H}_{G\setminus e} = n$, and $\hat{H}(G) = n/(n-D/2)$.
    Finally, let $G$  be a ``sunrise'', that is a two-loop graph with three edges. 
    Cutting each of its edge produce a one-loop graph, so that $H_G = 3 \times \frac{ 2}{2-D/2}$.
\end{example}

We will give a slightly different proof of \eqref{eq:borinsky}, which easily generalizes to the case of curve integrals.
We begin by proving a handy formula,
\begin{align}
    H_G = \sum_{e \in \mathrm{Edges}(G)}\int_0^1 d\alpha_{e'}\ (\mathcal{U}_{G \setminus e}^{\rm tr})^{-\frac{D}{2}},
    \label{eq:formula1}
\end{align}
where we have collectively denoted with $\alpha_{e'}$ the variables different than $\alpha_e$.
To prove \eqref{eq:formula1}, we decompose the domain of integration,
\begin{align*}
\mathbb{P}^{E_G-1}_{\ge 0} = \{\alpha_e \ge 0, e \in \mathrm{Edges}(G)\}/\mathbb{R}_{> 0},
\end{align*} into smaller domains, 
\begin{align*}
\mathcal{D}_e = \{\alpha_e \ge \alpha_{e'}\}/\mathbb{R}_{> 0} = \{1 \ge \alpha_{e'}\}.
\end{align*}
In these regions the maxima appearing in the tropical integrand can be evaluated explicitly. For this we make use of the cut/deletion property \eqref{eq:cutdeletion}, and note that in $\mathcal{D}_e$ each of the monomials appearing in $\mathcal{U}_{G/e}$ is bounded from above by some monomial appearing in $\alpha_e \mathcal{U}_{G\setminus e}$. This concludes the proof.

 The above brings us to a formula for $\hat{H}_G$,
 \begin{align}
     \hat{H}_G = \int_0^1 d\alpha_e \ (\mathcal{U}^{\mathrm{tr}}_G)^{-\frac{D}{2}}.
     \label{eq:formula2}
 \end{align}
The proof is similar to the one just discussed.
We begin by decomposing the cube that forms the domain of integration, $\{0\le \alpha_e \le 1 \}$, into the domains $\mathcal{D}_e = \{0 \le \alpha_e' \le \alpha_e \le 1 \}$ (sligthly abusing the notation of the previous proof). 
We re-parametrize the domain $\mathcal{D}_e$ by rescaling $\alpha_{e'} \to \alpha_e \alpha_{e'}$
The analysis depend on whether the edge $e$ carries a loop momentum or not. In the former case, in the region $\mathcal{D}_e$ it holds $\mathcal{U}^{\mathrm{tr}}_G = \alpha_e \mathcal{U}_{G\setminus e}^{\mathrm{tr}}$,  and furthermore $L_{G} = L_{G\setminus e} + 1$. In the latter case cutting $e$ produces a factorized graph $G\setminus e = G_L \times G_R$, and $\mathcal{U}_G = \mathcal{U}_{G_L} \times \mathcal{U}_{G_R}$ factorizes accordingly. In either case, after rescaling we have
\begin{align}
    \int_0^1 d\alpha_e\ (\mathcal{U}^{\mathrm{tr}}_G)^{-\frac{D}{2}} &= \sum_{e}\int_0^1 d\alpha_e \ \alpha_e^{d_{G}-1} \times \int_{0}^1 d\alpha_{e'}\ (\mathcal{U}^{\mathrm{tr}}_{G\setminus e})^{-\frac{D}{2}} \\
    & = \frac{1}{d_G}\sum_{e} \int_{0}^1 d\alpha_{e'}\ (\mathcal{U}^{\mathrm{tr}}_{G\setminus e})^{-\frac{D}{2}} = \hat{H}_G,
\end{align}
as desired.

A consequence of the last formula is that if $G$ has two disconnected components $G_L, G_R$, so that $\mathcal{U}_G = \mathcal{U}_{G_L} \times \mathcal{U}_{G_R}$, then $\hat{H}_G =\hat{H}_{G_L}\times \hat{H}_{G_R}$, even though neither $H_G$ nor $d_G$ factorize individually:
\begin{align}
    \hat{H}_G = \int_0^1 d\alpha_e \ (\mathcal{U}^{\mathrm{tr}}_G)^{-\frac{D}{2}} = \int_0^1 d\alpha_L (\mathcal{U}^{\mathrm{tr}}_{G_L})^{-\frac{D}{2}} \times \int_0^1 d\alpha_R (\mathcal{U}^{\mathrm{tr}}_{G_R})^{-\frac{D}{2}}  = \hat{H}_{G_L} \times \hat{H}_{G_R}.
    \label{eq:heppFact}
\end{align}
It follows immediately that
\begin{align}
    H_{G_L \times G_R} = \frac{d_L+d_R}{d_L d_R} H_L \times H_R,
\end{align}
which should be compared with \eqref{eq:almostfactgraphs}.
Surprisingly, the same argument shows that $\hat{H}_G$ factorizes whenever $\mathcal{U}_G$ factorizes, even if the graph $G$ is connected!
\begin{example}
    Let $G$ be a tree with $E$ edges. Since $\mathcal{U}_G = 1$ factorizes into the product of the $E$ one-edge subgraphs $\gamma_e$ of $G$, we have that $H_G = (H_1)^E = 1^E = 1$.
    Now let $G$ be ``buquet of tadpoles'', i.e. a graph with $E$ edges all adjacent to a single vertex. The $\mathcal{U}_G = \prod_{e \in \mathrm{Edges}(G)} \alpha_e$, from which again $H_G = 1$.   
\end{example}
\begin{example}
 Consider again the chain of bubble diagram of Figure \ref{fig:bubbles}.
 We have
\begin{align}
\mathcal{U}_G= \mathcal{U}_{\gamma_{1,2}} \times \mathcal{U}_{\gamma_{3,4}} =(\alpha_1+\alpha_2)\times(\alpha_3+\alpha_4).
\end{align}
Its Hepp bound is then (fixing $\alpha_5=1$)
\begin{align}
    \hat{H}_G &= \frac{1}{3}\int_{\mathbb{R}^4_{\ge 0}} \frac{d\alpha_1d\alpha_2d\alpha_3d\alpha_4}{\max(\max(\alpha_1,\alpha_2),\max(\alpha_3,\alpha_4))^{}\max(1,\alpha_1, \dots,\alpha_4)^{3}} = \\ &= 2\times2 = \hat{H}_{\gamma_{1,2}}\times\hat{H}_{\gamma_{3,4}},
\end{align}
in agreement with \eqref{eq:heppFact}.
\end{example}

Finally, putting together \eqref{eq:formula1} and \eqref{eq:formula2} we immediately prove \eqref{eq:borinsky}.

\subsection{Entire Amplitudes}

We define the \emph{surface Hepp bound} to be the curve integral
\begin{align}
    H_S &= \int_{\mathbb{P}^{E_S-1}} \frac{\omega}{\mathrm{GL}(1) \times \mathrm{MCG}_S} \mathcal{I}_S^{\rm tr},
    \label{eq:surfHepp}
\end{align}
associated to the tropical integrand
\begin{align}
    \mathcal{I}^{\rm tr}_S \coloneqq \symUtr_S(\alpha_C)^{d_S-\frac{D}{2}}\ \symFtr_S(\alpha_C)^{-d_S}.
    \label{eq:trint}
\end{align}
The tropical surface polynomials are
\begin{align}
    \mathcal{U}_S^{\mathrm{tr}} = \max_{\mathcal{C} \in \mathrm{Cuts}_1(S)}\prod_{C \in \mathcal{C}} \alpha_C,
\end{align}
\begin{align}
    \mathcal{F}_S^{\mathrm{tr}} \coloneqq \max\left(\max_{\mathcal{C} \in \mathrm{Cuts}_2(S)}\prod_{C \in \mathcal{C}} \alpha_C,\max_{\substack{C \in \mathrm{Curves}(S)\\ 
    \mathcal{C}' \in \mathrm{Cuts}_1(S)}}\alpha_C \prod_{C' \in \mathcal{C}'} \alpha_{C'} \right).
\end{align}
Note that there are two levels of tropicalization involved. One is when we think of the surface polynomials as polynomials in the variables $\alpha_C$, and we tropicalize them. And the second is interpreting $\alpha_C$ \emph{themselves}, as tropical functions.
This last level plays little role in this paper. All that will be necessary to know is that if $C$ and $C'$ are intersecting curves on a surface, then $\alpha_C \times \alpha_C'$ is a vanishing function on the Feynman fan.
The first level, instead, is crucial\footnote{ 
This begs to define a proper notion of Newton polytopes for the surface Symanzik polynomials, and of the action of the MCG on them, but here we will content ourselves with some elementary observations.}.
To begin with, we observe that all monomials appearing in $\mathcal{F}_S^0$ are contained among those appearing in the mass term of $\mathcal{F}_S$, so that
\begin{align}
    \mathcal{F}^{\mathrm{tr}}_S = \left(\max_{C \in \mathrm{Curves}(S)} \alpha_C\right) \times \mathcal{U}^{\mathrm{tr}}_S,
\end{align}
so that we can rewrite the Hepp bound in a simpler way,
\begin{align}
    H_S =\int_{\mathbb{P}^{E_S-1}} \frac{\omega}{\mathrm{GL}(1) \times \mathrm{MCG}_S}
\ \left(\mathcal{U}^{\mathrm{tr}}_S\right)^{-\frac{D}{2}}\ \left( \max_{C \in \mathrm{Curves}(S)}  \alpha_C \right)^{-d_S}.
    \label{eq:surfHepp}
\end{align}

\begin{figure}
    \centering
    \includegraphics[width=0.5\linewidth]{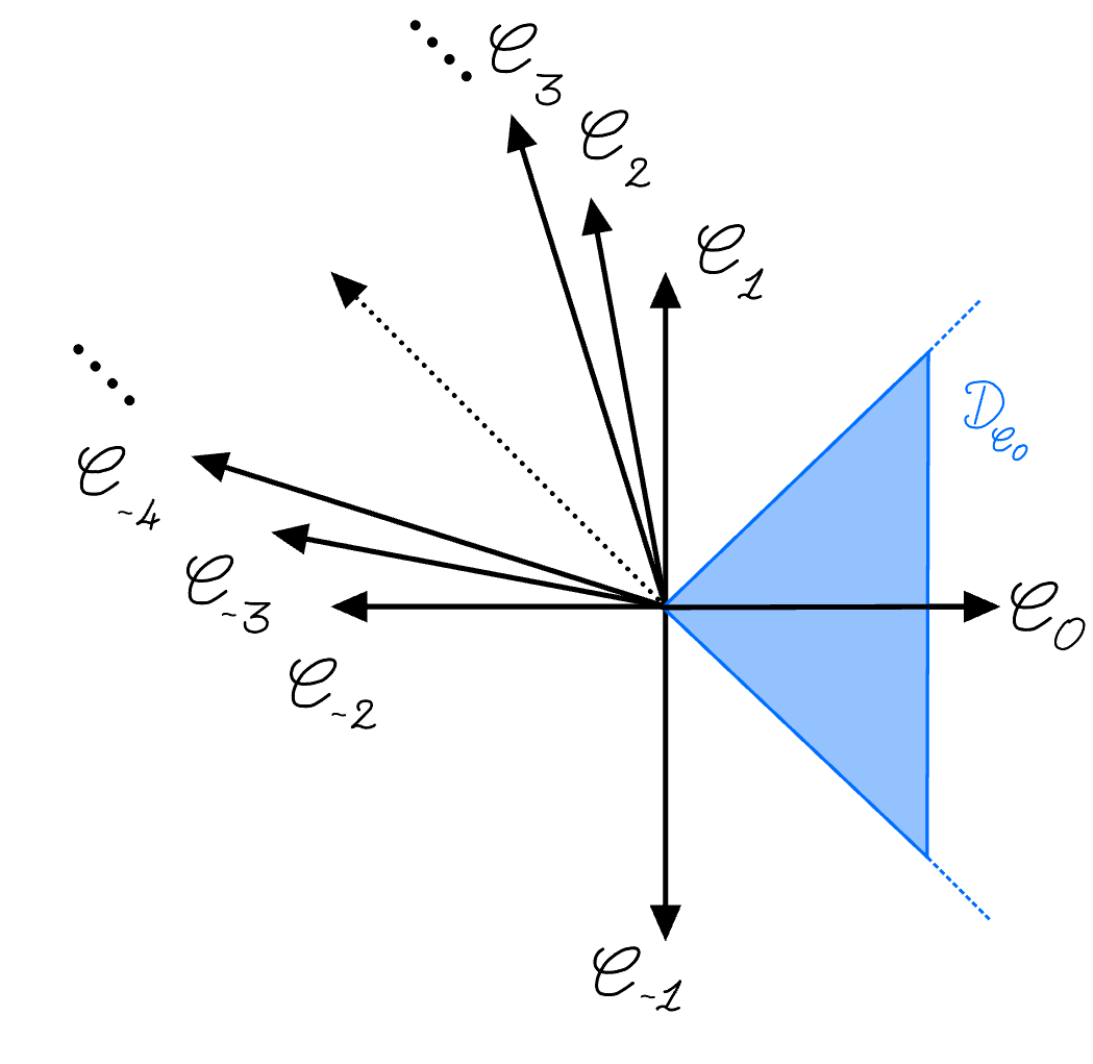}
    \caption{\emph{The dual fundamental domain.} A dual fundamental domain for the mapping class group is given by the cell $\mathcal{D}_C$}.
    \label{fig:a11bary}
\end{figure}

The surface Hepp bound can be evaluated explicitly by choosing a fundamental domain for the action of the MCG. 
One way to do so, is to enumerate the fatgraphs $\Gamma \in \mathrm{FatGraphs}(S)$, and for each to pick any cone $\Delta_\Gamma \in \Sigma_S$ in the Feynman fan $\Sigma_S$, which is labeled by it. The integral over $\Delta_\Gamma$ verbatim reproduces the Hepp bound of the graph $\Gamma$, so we get
\begin{align}
    H_S = \sum_{\Gamma \in \mathrm{FatGraphs}(S)} H_\Gamma.
\end{align}
\begin{example}
    Let $S = \{ 0, \{4\}\}$  be a disk with four marked points on the boundary.
    There are exactly two diagrams that can be drawn on $S$, these are tree graphs so $H_S = 1 + 1 = 2$.
\end{example}
\begin{example}
    Next consider $S = \{0, \{1,1\}\}$. Up to mapping class group there is only one graph that can be drawn on $S$, it is a one-loop $2$-gon, so $H_S = 2$.
\end{example}
In addition to the computational complexity of evaluating Hepp bounds of graphs, we now have to sum over a large number of graphs.
The rest of this section will be dedicated to illustrate a more efficient formula for the Hepp bound,
\begin{align}
    H_S = \sum_{C \in \mathrm{Curves}(S)/\mathrm{MCG}_S} \frac{H_{S\setminus C}}{d_{S\setminus C}},
    \label{eq:surfrec}
\end{align}
which involves a \emph{surface recursion}, evidently reminiscent of both \eqref{eq:Hepp2} and \cite{cutequation}.
\begin{example}
    \label{ex:treelevelsurfhepp}
    We continue the tree level example started in \ref{ex:treelevelcurveint}.
     The curve integral for the surface Hepp bound is 
    $$H_S=\int_{\mathbb{P}^{n-4}} \frac{\omega\ }{\mathrm{GL}(1)} \mathcal{I}^{\rm tr}_S = \int_{\mathbb{P}^{n-4}} \frac{\omega\ }{\mathrm{GL}(1)} \left( \max_{C \in \mathrm{Curves}(S)} \alpha_C  \right)^{n-3}\ . $$
    It can be calculated \emph{explictly},
    $$H_S = \sum_{G \in \mathrm{FatGraphs}(S)}\mathrm{Trop\ }\mathrm{Val}(G) = \sum_{\mathcal{C}=(C_1,\dots,C_{n-3}) \in \mathrm{Cuts}_{k}(S)}  (n-3) = (n-3)\times\mathrm{Catalan}(n-3),$$
    or \emph{recursively} as
    $$H_S = \sum_{C \in \mathrm{Curves}} \frac{H_{S\setminus C}}{E_{S\setminus{C}}},$$
    plus the boundary conditions 
    $$H_{\{0,\{3\}\}} = d_{\{0,\{3\}\}} = 1.$$
    By comparison with the \emph{cut equation} recursive representation of Catalan numbers \cite{cutequation}, we get $H_S =(n-3) \times \mathrm{Catalan}(n-3)$ and $\hat{H}_S = \mathrm{Catalan}(n-3).$
\end{example}
\begin{example}
Consider the surface $S=\{0,\{1,1\}\}$. Any curve on $S$ cuts the surface into a disk with four marked points on its boundary, so they are all equivalent up to MCG.
Therefore, the recursion gives
$H_S = \hat{H}_{\{0,\{4\}\}} = 2$
\end{example}
The key to proving \eqref{eq:surfrec} is to use a new decomposition of curve integrals, alternative to the usual sum over Feynman fatgraphs.
Consider the cone
\begin{align}
    \mathcal{D}_C = \{ t \in \mathbb{R}^{E_S}\ |\ \alpha_C(t) > \alpha_{C'}(t),\ \forall C' \in \mathrm{Curves}(S)\}.
    \label{eq:barycell}
\end{align}
The collection of cones $\mathcal{D}_C$ for $C \in \mathrm{Curves}(S)$ form a fan $\Sigma^*_S$, which is the barycentric dual of the Feynman fan $\Sigma_S$.
Furthermore, if $C$ runs over $\mathrm{Curves}(S)/\mathrm{MCG}_S$ then the collection of cells $\mathcal{D}_C$ forms a cell decomposition of a fundamental domain of the mapping class group, as shown in Figures \ref{fig:barycentric} and \ref{fig:a11bary}.
The proof then follows the same steps as those explained in the previous section, with obvious translations ``from edges of graphs'' to ``curves on surfaces''. 

Within each cell we fix the overall the scale by setting $\alpha_C = 1$. The cut/contraction property of Symanzik polynomials allows to evaluate the maxima appearing in the tropical integrand, which gives the formula
\begin{align}
    H_S = \sum_{C \in \mathrm{Curves}(S)/\mathrm{MCG}_S} \int_{1=\alpha_C\ge\alpha_{C'}} \frac{d\alpha_{C'}}{\mathrm{MCG}_{S \setminus C}}\ \left(\mathcal{U}^{\rm tr}_{S \setminus C}\right)^{-\frac{D}{2}}.
    \label{eq:surfHeppFormula1}
\end{align}
In the above, we have made use of the fact that the stabilizer of $C$, i.e. the subgroup of $\mathrm{MCG}_S$ that leaves invariate the curve $C$, is canonically isomorphic to the mapping class group of the surface $S' = S \setminus C$.

Next, we define $\hat{H}_S \coloneqq H_S/d_S$, and prove the formula
\begin{equation}
    \hat{H}_S = \int_{\alpha_C \le 1} \frac{\omega}{\mathrm{MCG}_S} \left(\mathcal{U}_S^{\rm tr}\right)^{-\frac{D}{2}}. \label{eq:surfHeppFormula2}
\end{equation}

Note that in \eqref{eq:surfHeppFormula2} we are implicitly defining a piecewise linear domain of integration in the Feynman fan as the locus where all headlight functions are less than one. 
This locus can be also described as the polytope obtained by taking the convex hull of the g-vectors $g_C$ associated to curves, i.e.
\begin{align}
    \mathcal{P}_S = \mathrm{Conv}\{ g_C, C \in \mathrm{Curves}(S)\}.
\end{align}
To the best of our knowledge, this polytope has not been studied before in the literature. 
En passant we remark that the polytope is triangulated by its intersection with the Feynman fan, which shows that its lattice volume (properly modded by the MCG) equals the number of Feynman diagrams. We warn that $\mathcal{P}_S$ is \emph{not} a dual ``surfacehedron''. For example, in the case of disks it is not a dual associahedron. In the conclusions, we will briefly comment of a possible application of $\mathcal{P}_S$ for a different sampling algorithm.

As usual, to prove \eqref{eq:surfHeppFormula2} we intersect the domain of integration $\mathcal{P}_S/\mathrm{MCG}_S$ with the barycentric decomposition.
In each cell, we rescale the headlight functions by the dominant one, integrate the overall scale and resolve the maxima of the tropical integrand by the cut/deletion property
\begin{align}
    \mathrm{RHS\ of\ } \ref{eq:surfHeppFormula2}
    &=\sum_{C \in \mathrm{Curves}(S)/\mathrm{MCG}_S} \int_0^1 d\alpha_C \alpha_C^{d_{S\setminus C}-1} \times\int_{\alpha_{C'} \le 1} d\alpha_C' \left(\mathcal{U}_{S\setminus C}\right)^{-\frac{D}{2}} \\
    &= \frac{1}{d_{S\setminus C}} \sum_{C \in \mathrm{Curves}(S)/\mathrm{MCG}_S} \int_{\alpha_{C'} \le 1} d\alpha_C' \left(\mathcal{U}_{S\setminus C}\right)^{-\frac{D}{2}}  = \hat{H}_S.
\end{align}
We have also used the fact that if $C$ is a factorizing curve, then $\mathcal{U}_{S}|_{\mathcal{D}_C} = \mathcal{U}_{S_L} \times \mathcal{U}_{S_R}.$

In order to complete the proof of the recursion, we would like to put together \eqref{eq:surfHeppFormula1} and \eqref{eq:surfHeppFormula2}. This is where our construction crucially makes use of a basic property of the Feynman fan. Imagine of projecting through a vector $g_C$ that part of the fan $\Sigma_S$ which is compatible with $g_C$\footnote{That is the cones that have $g_C$ among their rays.}. We claim that the projection coincides with the simpler Feynman fan $\Sigma_{S\setminus C}$ (in appropriate coordinates). As it was illustrated at length in \cite{counting1}, this geometrical statement is what encodes the unitarity of surface ordered amplitudes. 
In our setting, it concretely boils down to recognizing that the domain $\{1=\alpha_C \ge \alpha_{C'}, \forall C' \in \mathrm{Curves}(S) \}$ appearing on the RHS of \eqref{eq:surfHeppFormula1} coincides with the polytope $\mathcal{P}_{S\setminus C}$ that forms the integration domain of the RHS of \eqref{eq:surfHeppFormula2}.

With the benefit of hindsight, let us note that $\hat{H}_{S_L \times S_R} = \hat{H}_{S_L} \times \hat{H}_{S_R}$ again factorizes on disconnected surfaces, from which it follows immediately 
\begin{align}
    H_{S_L \times S_R} = \frac{d_{S_L}+d_{S_R}}{d_{S_L} d_{S_R}} H_{S_L} \times H_{S_R}.
\end{align}

\section{Sampling The Feynman Fan}
\label{sec:sampling}

We are now equipped with all the tools necessary to address the central problem of this paper: the numerical evaluation of curve integrals by a Monte Carlo approach.

The strategy is straightforward.
The first step is to choose a probability distribution function $f$ over the Feynman fan, $f(t)$, normalized so that $\int_{\mathbb{P}^{E_S-1}/\rm{MCG}_S} dt \ f(t) = 1$. 
Then one samples $N$ points $t_i$ distributed according to it $f$ and evaluates the average
\begin{align}
    A^{N}_S  \coloneqq \frac{1}{N}\sum_{i=1}^N \frac{\mathcal{I}_S(t_i)}{f(t_i)},
\end{align}
where $\mathcal{I}_S$ is the curve integrand. 
By the laws of large numbers, this estimate converges (in a statistical sense) to the true value as the number of sample increases, 
\begin{align}
    A^{N}_S \to A_S \quad \mathrm{as} \quad N\to\infty.
\end{align}
The precision of the estimate is measured by the variance of the random variable $A^{N}_S$, which is
\begin{align}
    \mathrm{Var}(A^{N}_S) = \frac{\mathrm{Var}_f(\mathcal{I}_S/f)}{\sqrt{N}}.
    \label{eq:mcaccuracy}
\end{align}
Increasing the number of samples is therefore the simplest way to obtain better estimates.

Another possiblity is to look for a sampling distribution such that the variance $\mathrm{Var}_f(\mathcal{I}_S/f)$ is smaller, an idea known as \emph{importance sampling}.
There are numerous excellent general-purpose algorithms that implement this idea.
In our case, however, it is reasonable to expect that the best results will be obtained if the geometrical structure of curve integrals is taken into account, rather than by using a generic strategy.

In \cite{Borinsky:2020rqs}, it was proposed that the tropical integrand is a natural candidate for a sampling distribution.
This was motivated by the observation that the tropical integrand bounds the original integrand from above and below. As already anticipated in the Introduction, it is natural to extend this idea to the context of curve integrals, and to use $\mathcal{I}_S^{\rm tr}$, as defined above, as a sampling distribution for curve integrals.

With this idea in mind, what is left to do is to find a way to generate samples distributed according to $\mathcal{I}_S^{\rm tr}$. 
In order to obtain a good estimate for $A_S$ we need to generate many samples. 
The main bottleneck of our Monte Carlo strategy is therefore the amount of time that is required to generate a single sample.
We will begin by discussing an algorithm that naively scales extremely well, requiring only $\mathcal{O}(1)$ time to generate a sample point.

\subsection{Feynman Sampling}

Consider a surface $S$, we define its \emph{sectors} to be ordered $E_S$-tuples of its curves,
\begin{align}
    \mathrm{Sectors}(S) = \{(C_1, \dots, C_{E_S}), C_i \in \mathrm{Curves}(S)\}.
\end{align}
From now on, we will use round brackets to denote ordered sets.
The action of the MCG extends to sectors in the obvious way. A sector of $S$ up to MCG is identified by a fatgraph of $S$, together with an ordering of its edges.
It follows that the number of sectors is very large, $|\mathrm{Sectors}(S)|=E_S! \times |\mathrm{FatGraphs}(S)|$.

A sector $\mathcal{S}=(C_1,\dots,C_{E_S})$ defines a region in the Feynman fan,
\begin{align}
    \Delta_{\mathcal{S}} = \{\alpha_{C_i}>\alpha_{C_{i+1}}, i=1,\dots,E_S\}.
\end{align}
The barycentric cell $\mathcal{D}_C$, that we have introduced in Section \ref{sec:tropicalization}, is triangulated by the union of all simplices of the form $\Delta_{(C, \dots)}$.
Any collection of simplices $\Delta_\mathcal{S}$, with $\mathcal{S}$ ranging over $\mathrm{Sectors}(S)/\mathrm{MCG}_S$, defines a fundamental domain for the action of the MCG on the Feynman fan.
However, it may not always be possible to recompose these sectors into a Feynman fundamental domain, as in Figure \ref{fig:a11bary}. 

For any sector $\mathcal{S} = (C_1,\dots,C_{E_S})$, we define a monomial parametrization of $\Delta_\mathcal{S}$, by
\begin{align}
    \alpha_{C_j} = \prod_{i=1}^{j}  t_i,
    \label{eq:alphasoft}
\end{align}
with $t_i = \in [0,1]$.
Note that in these coordinates the simplex $\Delta_\mathcal{S}$ looks like a cube.
The Hepp bound curve integrand is particularly simple when expressed in terms of the cubical parametrization,
\begin{align}
    \left. \frac{\omega}{\mathrm{GL}(1)\times\mathrm{MCG}_S}\ \mathcal{I}_S^\mathrm{tr}\right|_{\mathcal{S}} = \prod_{i=2}^{E_S} dt_i\ t_i^{d_{S\setminus C_1 \dots C_{i-1}}-1},
\end{align}
we have fixed the $\mathrm{GL}(1)$ redundancy by imposing $\alpha_1 =t_1 =1$, while the $\mathrm{MCG}$ is completely stabilized in a sector. 
It follows immediately that the contribution $H_\mathcal{S}$ coming from this sector to the total Hepp bound is 
\begin{align}
H_\mathcal{S} = \prod_{i=1}^{E_S} \frac{1}{d_{S\setminus C_1 \dots C_i}},
\end{align}
which is of course equivalent to \eqref{eq:HeppSec}.
This suggests to further reparametrize the cube by rescaling $\log t_i$ by $d_{S\setminus C_1 \dots C_{i}}$. In other words, we change variables to 
\begin{align}
    t_i = \hat{t}_i^{\frac{1}{d_{S\setminus C_1 \dots C_i}}}.
    \label{eq:that}
\end{align}
The tropical integrand is a uniform distribution over the hypercube parametrized by the new coordinates $\hat{t}$.

We are now ready to formulate the first algorithm to sample a fundamental domain of the Feynman fan according to the tropical curve integrand. 
First, we enumerate a copy of all sectors, up to MCG, and for each sector we compute the probability $$p(\mathcal{S}) = \frac{H_\mathcal{S}}{H_S}.$$ 
This is a pre-processing step which has to be done only once. Having computed and stored these data, we can define the sampling algorithm proper.
First, we sample uniformly a point $\hat{t}$ in the hypercube $[0,1]^{E_S}$, then we draw a sector according to $p(\mathcal{S})$, and then return the point with coordinates $\hat{t}$ in $\Delta_\mathcal{S}$.

Let us discuss the computational complexity of this sampling algorithm.
Using the alias method it is possible to draw from a discrete distribution with $N$ possible outcomes in $\mathcal{O}(1)$ time, regardless of the value of $N$. 
Naively this seems to suggest that the best way to evaluate curve integrals is to go back to enumerating sectors which are, after all, Feynman diagrams!
However, the alias method requires storing the $N$ probabilities defining the distribution, so that requires $\mathcal{O}(N)$ amount of space (that is memory on a computer).
Considering the large number of sectors on a surface, this algorithm becomes quickly unfeasible.

A better Feynman-based algorithm, would be to sample diagrams rather than sectors. Having picked a diagram, one could then apply Algorithm 4 from \cite{Borinsky:2020rqs} to sample within its simplex. 
This gets rid of the factorial growth of sectors, but does nothing to address the growth of the number of diagrams, which again renders the algorithm unfeasible very quickly. 

We will next discuss an alternative sampling procedure, which trades a small amount of time for a drastic amount of space. In particular, we will see that a crucial aspect of the new algorithm is that it does not require to generate and store Feynman diagrams.
The result is a sampling procedure that it is feasible long after the naive algorithm above has ceased to be.

\subsection{Dual Sampling}

In order to address the limitations of the previous algorithm, we now follow a different path, inspired by the permutohedral sampling algorithm from \cite{Borinsky:2020rqs}.

We will consider a stochastic process that at each step draws from the set of curves on a surface $S$. After $E_S$ steps we have a sequence of curves $(C_1, \dots,C_{E_S})$, which defines a sector $\mathcal{S}$ of $S$. Therefore the stochastic process gives a sampling over sectors. 
Furthermore, we will prove shortly that a sector $\mathcal{S}$ is picked with probability $p(\mathcal{S})=\frac{H_{\mathcal{S}}}{H_S}$. It follows that the stochastic process can be used to sample the Feynman fan according to the tropical integrand $\mathcal{I}_S^{\rm tr}$.
Compared with the previous algorithm, the preparation step is much more efficient in terms of space. This is  because we have to store probabilities associated to curves, and curves are much fewer than either sectors or diagrams.

Let us begin by describing the sampling process in plain terms, and then discuss the complexity aspects of its implementation.
Let $S$ be a surface, which may be disconnected, and label all equivalence classes of curves up to MCG by how they cut $S$.
For each curve $C$, compute and store the probability 
\begin{align}
    p(C|S)=\frac{\hat{H}_{S\setminus C}}{H_S}
    \label{eq:probCurveGivenSurface}
\end{align}
Having dealt with this preparation step, the sampling proper can be implemented as follows.
Given a surface $S$, pick a curve $C \in \mathrm{Curves}(S)/\mathrm{MCG}_S$ with probability $p(C|S)$. Iterate the sampling with $S$ replaced by $S \setminus C$. 
After $E_S$ steps, the probability for a sector $\mathcal{S} = (C_1,\dots,C_{E_S})$ to be produced is 
\begin{align*}
 p(\mathcal{S}) &= p(C_1|S)p(C_2|S\setminus C_1)\dots p(C_{E_S}|S\setminus C_1\dots C_{E_{S}-1}) \\ 
 &= \frac{\hat{H}_{S\setminus C_1}}{H_S} \frac{\hat{H}_{S\setminus C_1 C_2}}{H_{S\setminus C_1}}\dots \frac{\hat{H}_{S\setminus C_1\dots C_{E_S}}}{H_{S\setminus C_1\dots C_{E_S-1}}} 
 = \frac{\frac{\cancel{H_{S\setminus C_1}}}{d_{S\setminus C_1}}}{H_S} \frac{\frac{H_{S\setminus C_1 C_2}}{d_{S\setminus C_1 C_2}}}{\cancel{H_{S\setminus C_1}}} \dots \frac{1}{H_{S\setminus C_1\dots C_{E_S-1}}}
 = \frac{\frac{1}{d_{S\setminus C_1}d_{S\setminus C_1 C_2}\ \dots\ 1}}{H_S}.
\end{align*}
Therefore the sampling over sectors agrees with the direct algorithm of the previous section.

The algorithm can be drastically improved by taking advantage of the following observation. Suppose that at some stage of the sampling process, a factorized surface of the form $S_L \times S_R$ is produced.
The probability to pick, say, the curve $C_L$ on $S_L$ at the next step is given by
\begin{align}
    p(C_L|S_L\times S_R) &=\frac{\hat{H}_{S_L\setminus C_L}  \hat{H}_{S_R}}{H_{S_L \times S_R}}  =\frac{\hat{H}_{S_L\setminus C_L}  \cancel{\hat{H}_{S_R}}}{(d_{S_L}+d_{S_R}) \hat{H}_{S_L} \cancel{\hat{H}_{S_R}}} = \frac{d_L} {d_{S_L}+d_{S_R}} \frac{\hat{H}_{S_L\setminus C_L} }{H_{S_L}} \nonumber\\
    &= \frac{d_L}{d_L+d_R} p(C_L|S_L) = p(S_L | S_L \times S_R) p(C_L|S_L).
    \label{eq:factinconnected}
\end{align}
This shows that the process \emph{factorizes} into two steps: picking a connected component of $S_L \times S_R$, with probability 
\begin{align}
    p(S_L | S_L \times S_R)\coloneqq\frac{d_L}{d_L + d_R},
    \label{eq:sampleconnectprob}
\end{align}
and then picking curves on this component by the usual rule for connected surfaces.
Note that from the point of view of a connected component, curves will be drawn on it ever since by a stochastic process which is \emph{independent} from those of the other components.
However, the times at which curves are picked on different connected components, which gives the relative ordering of the corresponding headlight variables $\alpha_C$, are not independent from each other \footnote{Because of this, the process does not factorize into perfectly independent sub-processes. It would be interesting to explore how to exploit this imperfect factorization to increase performance by a parallelization of the sampling algorithm.}.
The bottomline is that we have to compute and store Hepp bounds associated to connected surfaces only.

For the benefit of clarity, the \emph{Dual Sampling Algorithm} is summarized in the pseudo-code below.
\begin{algorithm}[H]
\begin{algorithmic}
\State $\textbf{input:}$ A surface $S$ $\textbf{output:}$ A sector of $S$
\State{Set $i \gets 1$}
\State{Set $S_{(i)} \gets S$}
\While{$i < E_S+1$}
\State{Sample a connected component $S^{\rm conn}$ of $S$ with probability $p(S^{\rm conn}|S) = \frac{d_{S^{\rm conn}}}{d_S}$}
\State {Sample a curve $C_{i}$ on $S^{\rm conn}$ with probability $p(S'=S^{\rm conn}\setminus C|S^{\rm conn})=\frac{\hat{H}_{S'}}{H_{S^{\rm conn}}}$}
\State{Set $S_{(i+1)}$ to be $S_{(i)}$ with $S^{\rm conn}$ replaced by $S'$}
\State{Set $i \gets i+1$}
\EndWhile{}
\State{\textbf{return} $\mathcal{S}=(C_1,\dots,C_{E_S})$}
\end{algorithmic}
\caption{Dual Sampling Algorithm}\label{alg:dualsampler}
\end{algorithm}

Let us now address the complexity of the dual sampling algorithm.
It is important to keep in mind the tension between the time efficiency, $\mathcal{O}(1)$ time to draw from a discrete probability regardless of the number of possible outcomes, and space efficiency, the number of Hepp bounds that have to be precomputed and stored.
The dual sampling algorithm requires drawing curves from surface a total of $E_S$ times, resulting in a $\mathcal{O}(E_S)$ time.
In terms of space, we need to generate Hepp bounds $\hat{H}_{S \setminus C}$ for various surfaces $S$ and all curves up to MCG on $S$.
Thanks to the previous observation, we can restrict to connected surfaces only. Furthermore, recall that Hepp bounds do not depend on the labels of boundary segments, reducing the total number of unique bounds to compute and store.

As a simple example, consider the case of a disk with $n$ marked points. Recall that we can compute the curve integral directly, with no need of a Monte Carlo numerical integration, but this examples is useful to illustrate the space efficiency of the dual algorithm.
If we were to sample over the sectors by the algorithm of the previous subsection, we would have to generate the all the cubic planar diagrams, the number of which is the $n$-th Catalan number, that asymptotically grows exponentially.
On the other hand, the dual algorithm requires to store all Hepp bounds for all connected unlabelled disks with $n' \le n$ marks. These were calculated in Example \ref{ex:treelevelsurfhepp}, essentially they coincide with Catalan numbers. So, rather than enumerating exponentially many diagrams, we calculate and store all Catalan numbers up to the $n$-th.

There is also a pleasant surprise, which only becomes evident during runtime.
Suppose that, in the dual sampling algorithm, we did not separate the pre-processing step from the sampling proper.
We could compute surface Hepp bounds only \emph{when needed} and store them for later usage,
This way after obtaining a certain number of samples, very unlikely stories of the stochastic decision tree will not have been explored and consequently stored in memory.

In conclusion, the dual algorithm allows to sample the Feynman fan, according to the Hepp curve integrand, reaching surfaces with much higher loop number than the direct sampling by Feynman diagrams can reach.

\subsection{Why ``Dual''?}

Both the algorithm of \cite{Borinsky:2020rqs} and the dual algorithm explained here, work by organizing the data of sectors in a more compact container: the decision tree of a stochastic process.

In fact, the Feynman sampling described above, which prepends the sampling of a diagram to Algorithm 4 of \cite{Borinsky:2020rqs}, and the dual sampling are realized by stochastic processes that traverse the same decision tree, but in \emph{opposite directions}.

\begin{figure}
    \centering
    \includegraphics[width=0.35\linewidth,angle=-90]{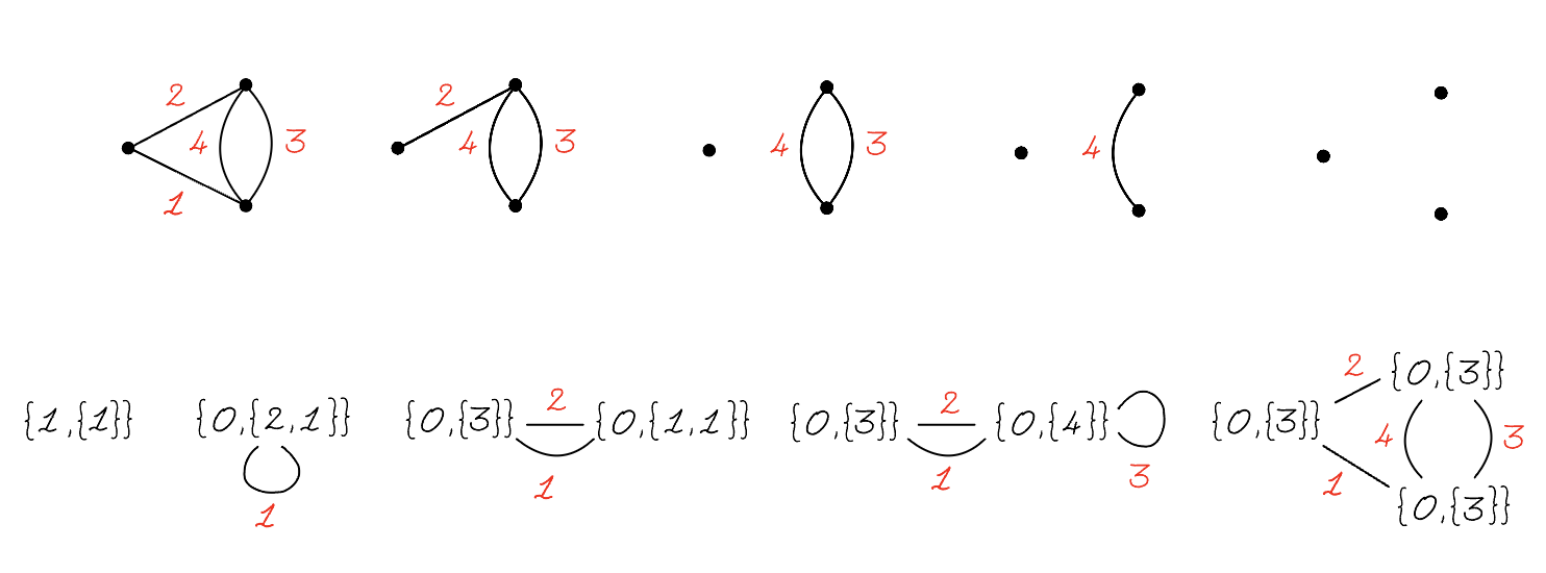}
    \caption{\emph{Dual processes.} The stochastic process of \cite{Borinsky:2020rqs} starts from a graph and \emph{removes} its edges (Top).
    The dual process starts from a vertex labelled by a surface and \emph{adds} edges (Bottom).
    }
    \label{fig:dualproc}
\end{figure}

This is vividly seen when we look at a trajectory of each process, as shown in Figure \ref{fig:dualproc}.
Algorithm 4 of \cite{Borinsky:2020rqs} starts from a diagram, randomly removes an edge from it and then iterates on the new graph.
On the other hand, the dual algorithm starts from a vertex --- labeled by a surface --- and recursively splits it into smaller vertices connected by edges. 

In particular, in the dual algorithm the Feynman fatgraph is not chosen from the start, but rather gradually built as the process evolves. This is why it is not necessary to explicitly enumerate the fatgraphs to run the dual algorithm.

\subsection{Implementation}

This paper comes with an ancillary file, a \verb|Mathematica| notebook providing an implementation of the dual sampling algorithm.
Here we give a brief description of its contents.

The implementation consists of three main modules:
\begin{itemize}
    \item Surface Recursion
    \item Dual Tropical Sampling
    \item Parametric Curve Integrand
\end{itemize}

We denote labeled connected surfaces by \verb|csurf[g,part]|, giving genus and the partition of the boundary segments into sub-lists. Each sub-list should be treated up to cyclic order and the partition should not be treated as ordered. For each boundary segment we must provide the data $\verb|z|[\mathrm{ext},\mathrm{label},\mathrm{homology}]$: it declares the curve as an external curve, with a unique label and an homology (the momentum of the external particle on the boundary segment).
Disconnected surfaces are un-ordered lists of connected surfaces.
We also denote unlabeled surfaces by $\verb|S[g,nmarks]|$, giving genus and the list of the number of marks on each boundary.
At this stage only un-punctured surfaces are implemented, but an extension to more general surfaces is straightforward.

{\bf Surface Recursion.} The function $\verb|hepp[S]|$ returns the surface Hepp bound for the surface $S$. It is calculated by the recursion \eqref{eq:surfrec}. 
The code also implement the surface recursion to compute the number of diagrams via the function \verb|numberDiags[S]|. 
To achieve a greater speed, the recursion is implemented at the level of unlabeled surfaces directly.

{\bf Dual Sampling.} 
The first important function is \verb|cutConnectedSurface[S]|, which returns all the curves up to MCG on the connected surface \verb|S|. Each class is labeled by the possibly disconnected surface produced cutting along a curve representative.
This produces a new boundary segment, which is automatically given a unique label (also recording the time at which the curve was drawn during the stochastic process) and declared either as a factorizing curve or loop curve. 
Its homology is accordingly either computed in terms of previously picked curves or declared as a new loop variable.
Calculating the data of the homology already at this step is very useful because it will speed up the evaluation of the integrand at the sampled point.
The helper functions $\verb|toCanonicalSurface|$ and $\verb|fromCanonicalSurface|$ allow to pass back and forth between a surface with specific boundary segments labels and one with standard labels. This is what allows us to store Hepp bounds only for unlabeled surfaces, saving considerable amount of space for a negligible quantity of time: we sample curves from canonical surfaces and then restore the mark data of the actual surface.

With these functions it is easy to implement the stochastic process on curves.
The pre-processing step is dealt with by the function $\verb|getCutRules|$. 
Applied to a \emph{connected} surface $S$ it returns the rule \begin{align}
    \{p(C|S), C \in \mathrm{Curves}(S)/\mathrm{MCG}_S\} \to \{S'=S\setminus C, C \in \mathrm{Curves}(S)/\mathrm{MCG}_S\},
\end{align}
which can be used to draw curves up to MCG  on $S$ using the built-in function \\
\verb|RandomChoice|.
The function also acts as a container, since it remembers the values it has previously evaluated, storing them in the list $\verb|DownValues[getCutRules]|$.

The sampling is implemented through three functions.
The function \\ \verb|sampleSurface[S]| draws a connected component of the surface \verb|S| with probability \eqref{eq:sampleconnectprob}.
The function \verb|sampleCuts[S]| draws a curve $C$ on a \emph{connected} surface $S$ with probability \eqref{eq:probCurveGivenSurface}.
The function \verb|sampleSector[S]| draws a sector on $\verb|S|$, by iteratively sampling a connected component and then a curve on it.

A pleasant feature of this design, is that the pre-process step is actually done during the sampling procedure. 
Whenever new data is required, it is computed and saved for future use. This way, only a fraction of the tree describing all stories of the stochastic process on curves is actually generated: the very unlikely stories are never explored.
Accordingly, the user will notice that when evaluating amplitudes at higher loops or multiplicitly, the first samples will require a lot of time, this is because the sampler is computing the data for a lot of new stories encountered for the first time.
We provide an already trained version of the code, where the most likely stories up to ten loops have already been explored. It can be loaded by running the command $\verb|Get["decisionTree.mx"]|$.
In order to save computed data for later use, the user can write \verb|DumpSave[toCutRules,"newDecisionTree.mx"]| and then load it as usual.

{\bf Parametric Curve Integrand}
Finally, the helper function $\verb|evalIntegrand[sec]|$ evaluates the integrand at a random point of the sector \verb|sec|, distributed according to $\mathcal{I}^{\rm tr}$. 
The sector is presented as a disconnected surface formed by triangles glued by common sides.
Internally, the function draws uniformly from the cubical parametrization \eqref{eq:that} and computes its image under the monomial map \eqref{eq:alphasoft}.
The integrand is then numerically evaluated at this point. Note that it is far more efficient to compute the Symanzik polynomials from the graph Laplacian than from the formulae in terms of cuts and therefore we implement this method \cite{Borinsky:2020rqs}.
The tropical integrand, i.e. the sampling distribution, restricts in the sector to the product of the headlight variables $\alpha_C$ for the first $L_S$ \emph{non-factorizing} curves which have been picked.
Both integrands can therefore be easily evaluated from the boundary segment data that we have previously computed and stored in the representation of the sector.
We wish to stress that it is straightforward to modify \verb|evalIntegrand| to compute numerically \emph{any} Feynman parametric integrand, including theories beyond cubic scalars. 

With these function, the implementation of the dual sampling algorithm for the Monte Carlo integration is complete.

\subsection{Tests and Results}

We have used the provided code to evaluate amplitudes $A_S$ for various surfaces $S$, all evaluated at zero momentum. 

In order to validate our code, we have compared our results with the explicit formula in terms of fatgraphs, \eqref{eq:curveint0}.
For each surface $S$, we provide in the ancillary files an explicit list containing all the fatgraphs in $\mathrm{FatGraphs}(S)$. Each fatgraph is represented  as a list of its vertices. Each vertex is a cyclically ordered triple of integers. Each integer labels an oriented edge adjacent to the vertex.
For example, the surface $S = \{0,\{4\}\}$ contains two fatgraphs, $\Gamma_s =\{[1,2,5],[3,4,-5]\}$ and $\Gamma_t =\{[2,3,5],[4,1,-5]\}$.
For each fatgraph we have evaluated $I^{\rm Feyn}_\Gamma$ using \verb|feyntrop| \cite{Borinsky_2023,feyntrop}.
We report our results in Table \ref{tab:validation}. For all we find agreement with this procedure, within the estimated accuracy. The accuracy is estimated using the standard deviation of the sampled values.
\begin{table}[h]
    \centering
    \begin{tabular}{c | c | c | c | c | c | c}
        $S$ & $n$ & $L_S$ & $|\mathrm{FatGraph}(S)|$ & $n_{\rm samples}$   & $A_S$ & $\delta$ \\
        \hline
        \hline
        $\{0,\{1,1,1\}\}$ & 3 & 2  & 32 & $10^5$ & $14.8$ & $0.1$\\
        $\{1,\{3\}\}$ & 3 & 2 & 70 & $10^5$ & $23.5$ & $0.15$ \\
         $\{1,\{2,1\}\}$ & 3 & 3 & 1040 & $10^9$ & $157.60$ & $0.03$ \\
        $\{2,\{1\}\}$ & 1 & 4 & 105 & $10^5$ & $11.1$ & $0.16$ \\
        \hline
    \end{tabular}
    \caption{Validation results: number of particles, loop order, number of fatgraphs, number of samples, numerical result for the amplitude \eqref{eq:curveint0} at zero momentum and estimated accuracy. All results agree with those obtained using the computer program presented in \cite{Borinsky_2023}, within the reported accuracy.}
    \label{tab:validation}
\end{table}

We have explored the performance of our code as the loop order increases, and provide a benchmark in Table \ref{tab:results}. Samples have been obtained by running in parallel $100$ Mathematica kernels  on the \verb|juno| node of the IAS School of Natural Sciences computing cluster (Processor: 2x Intel Xeon E5-2680 v3 12-core, 2.5 GHz. Cores: 24). The results have then been averaged together to give our estimate for the amplitude.

\begin{table}[h]
    \centering
    \resizebox{\columnwidth}{!}{
    \begin{tabular}{c | c | c | c | c | c | c | c | c }
        $S$ & $E_S$ & $L_S$ & $|\mathrm{FatGraph}(S)|$ & $n_{\rm samples}$ & $t_{\mathrm{sample}} $ & $A_S$ & $\delta/A_S$ & Memory\\
        \hline
        \hline
        $\{1,\{1\}\}$ & $4$  & $2$ & $1$    & $10^2$ & $\mathcal{O}(10^{-4}\ \mathrm{s})$ & $7 \times 10^{-1}$  & $7\%$ & 3 KB \\
        $\{2,\{1\}\}$ & $10$ & $4$ & $105$  & $10^4$ & $\mathcal{O}(10^{-3}\ \mathrm{s})$ & $4 \times 10^{1}$   & $3\%$ & 43 KB \\
        $\{3,\{1\}\}$ & $16$ & $6$ & 50 050 & $10^4$ & $\mathcal{O}(10^{-3}\ \mathrm{s})$ & $6 \times 10^{2}$   & $5\%$ & 354 KB \\
        $\{4,\{1\}\}$ & $22$ & $8$ & 56 581 525 & $10^6$ & $\mathcal{O}(10^{-2}\ \mathrm{s})$ &  $7\times 10^{4}$ & $9\%$ & 2.2 MB \\
        $\{5,\{1\}\}$ & $28$ & $10$ & 117 123 756 750 & $10^7$ & $\mathcal{O}(10^{-2}\ \mathrm{s})$ & $1\times 10^{7}$ & $5\%$ & 11 MB \\
        \hline
    \end{tabular}
    }
    \caption{Results for surfaces of increasing genus: number of edges, loop order, number of fatgraphs, number of samples required to achieve target accuracy ($\le 10\%$), average time per sample (including the generation of cut rules and Hepp bounds), numerical result for the amplitude \eqref{eq:curveint0}, percentage relative accuracy, size of the final decision tree.
    All amplitudes listed here involve a single external particle with zero momentum.}
    \label{tab:results}
\end{table}

Looking at the data, we see that the time per samples grows mildly. This is expected: the sampling requires $\sim\mathcal{O}(E_S)\sim \# \mathrm{particles/loops}$ time to generate a sector and a similar polynomial amount of time goes in the evaluation of the integrand. 

The most interesting result reported in Table \ref{tab:results} is the number of samples required to achieve a certain target accuracy (here we content ourselves with reaching one significant digit, that is a relative accuracy $\le10\%$).
As we already anticipated in the introduction, what is intriguing is that we see that the number of samples required to achieve a fixed accuracy increases moderately with the complexity of the amplitude, in contrast with the number of diagrams.
In this regard it is important to stress that the number of \emph{graphs} contributing to $A_S$ is considerably smaller than the number of fatgraphs, as multiple fatgraphs may descend to the same graph. 
While we are not aware of any simple way to count the unique graphs that contribute to a given $A_S$, we can bound this number from above by counting all graphs in uncolored $\phi^3$ theory, and from below by sampling over fatgraphs and using the value of the polynomial $\mathcal{U}_G$ at $\alpha_C =1$ as a non-faithful invariant to recognize isomorphic graphs \footnote{If two graphs are isomorphich then they have the same value for $\mathcal{U}_G=1$, but the contrary is not true.}.
As an example, for the $10$-loop surface $A_S = \{5,\{1\}\}$ we estimate the number of unique graphs to be between $1 \times10^6$ and $6 \times 10^6$. 

Using \verb|feyntrop|, we have observed that the number of samples required to achieve the same target accuracy for the evaluation of an individual diagram contributing to $A_S$, is comparable to the numbers listed in Table \ref{tab:results}. For instance, in the \verb|Tests| directory of in the ancillary files, we report an example of one diagram contributing to $S=\{5,\{1\}\}$, which required $10^7$ samples to be evaluated with $\le 10\%$ accuracy using \verb|feyntrop|. In this example, the dual algorithm is then saving us from the expectation of having to generate $\sim 10^6 \times 10^7$ samples to evaluate $A_S$. This is the formidable gain that we have anticipated in the introduction.

\subsection{Connected Recursion}
\label{subsec:connectedrecursion}

Recall that the surface recursion \eqref{eq:surfrec} and the dual sampling that goes with it were derived from the barycentric decomposition of the Feynman fan given by cells
\begin{align}
    \mathcal{D}_C = \{\alpha_C > \alpha_{C'}, \forall C' \in \mathrm{Curves}(S)\},
\end{align}
with $C$ an arbitrary curve on the surface $S$.
However, other decompositions are possible, where we consider only a subset of all curves, that is
\begin{align}
    \mathcal{D}_C = \{\alpha_C > \alpha_{C'}, \forall C,C' \in \mathcal{S} \subseteq \mathrm{Curves}(S)\},
\end{align}
with $C$ in $\mathcal{S}$. Not all subsets can be used to define a decomposition: we have to guarantee that all diagrams have at least one curve in $\mathcal{S}$.
A particularly interesting example is the subset of non-factorizing curves. Recall that these are precisely those that carry a loop momentum.

Any viable choice of $\mathcal{S}$ results in a recursive formula for the Hepp bound, and a sampling algorithm. These formulae are more efficient than the one we considered here: at each step they involve fewer terms because only the curves in $\mathcal{S}$ contribute.
The price to pay is that the Hepp bound has to be partitioned in more contributions, each of which individually satisfy a recursion. This should be compared with the freedom in choosing a \emph{shift set} for the recursion of surface functions described in \cite{cutequation}.
For instance, when using only non factorizing curves, the recursion only involves connected surfaces. The Hepp bound is partitioned into the contributions coming from graphs with a fixed superficial degree of divergence (now calculated by rescaling only the propagators carrying a loop momentum). Accordingly, the overall factor $\Gamma(d_S)$ is replaced with a piecewise gamma function in the curve integrand.

Finally, these more efficient recursions can be used to design better dual sampling algorithms.
We will leave for a future work their detailed description and implementation.

\section{All Loop Scattering As An Optimization Problem}
\label{sec:samplingproblem}

We now describe a generalization of the dual sampling algorithm, which can be applied to more general scattering amplitudes.

Let us begin by retracing concisely the steps that brought us to the dual sampling algorithm.
Since we care about curves up to MCG, which are labeled by the surfaces produced upon cutting, it is convenient to phrase the construction in terms of surfaces only.

The first crucial result was to discover that the surface Hepp bound satisfies a \emph{surface recursion},
\begin{align}
    H_S = \sum_{S' \in \mathrm{Curves}(S)/\mathrm{MCG}_S} \frac{H_{S'}}{d_{S'}},
    \label{eq:recursion}
\end{align}
with \emph{recursion kernel} $d_S \coloneqq E_S - L_S$.
The sum runs over all surfaces $S'$ that can be produced cutting $S$ by a curve (up to MCG).

Secondly, we have defined a stochastic process which iteratively draws a random curve on a surface and cuts it.
The process is described by a random variable $S_{(i)}$ that at each time takes values from the set of possibly disconnected surfaces. 
Let $A_S$ be the amplitude we want to calculate.
The process begins with $S_{(1)} = S$.
At each step, the process draws a connected component $S^{\rm conn}$ of $S_{(i)}$, then draws one of the surfaces $S'$ that can be obtained by cutting $S^{\rm conn}$, and finally replaces $S^{\rm conn}$ by $S'$ producing a new surface $S_{(i+1)}$.
After $E_S$ steps the process terminates with $S_{(E_S+1)}$ being a collection of triangles, glued by their sides. A side in common to two triangles carries a label which records at which step of the process it was produced. Therefore $S_{(E_S+1)}$ defines a sector of $S$ up to MCG.
The stochastic process is completely characterized by the probabilities to jump from a surface to the next one, 
\begin{align}
    p(S_{(i)}\to S_{(i+1)})  &=p(S^{(i)}\to S^{\rm conn}) p(S^{\rm conn} \to S')
    \\&= \frac{d_{S^{conn}}}{d_{S_{(i)}}} \frac{H_{S'}/d_{S'}}{H_{S^{\rm conn}}}
    \label{eq:probs}
\end{align}

After a sector has been produced, we sample within any of the corresponding cones of the Feynman fan, using the cubical parametrization
\begin{align}
    \alpha_i = \prod_{j=1}^i \hat{t}_j^{\frac{1}{d_{S_{(j)}}}},
\end{align}
and sampling uniformly $t_j \in [0,1]$. The projective invariance is fixed by setting $\alpha_1 =t_1= 1$.

The dual sampling algorithm allows to generate points on the Feynman fan distributed according to the sampling distribution $\mathcal{J}_d \coloneqq\mathcal{I}_S^{\rm tr}/H_S$.
It is a piecewise monomial function, with respect to the refinement of the Feynman fan into sectors.
In the sector $\mathcal{S} = (C_1,\dots,C_{E_S})$, defined as the locus $\Delta_{\mathcal{S}}=\{0< \alpha_{C_i} \le \alpha_{C_{i+1}}\}$,
the distribution restricts to 
\begin{align}
    \left. \frac{\omega}{\mathrm{GL}(1)\times\mathrm{MCG}_S}\ \mathcal{J}_d\right|_{\mathcal{S}} = \frac{d\alpha}{H_S} \prod_{i=2}^{E_S} \alpha_i^{d_{ S_{(i)}}-d_{S_{(i+1)}}-1}.
    \label{eq:explicit}
\end{align}

We now propose a natural generalization of this entire construction, obtained by simply replacing the quantity $d_S = E_S - L_S$  with an \emph{arbitrary} recursion kernel 
\begin{align}
    d : \mathrm{Surfaces} \to \mathbb{R}_{\ge 0},
\end{align}
such that $d_{S_L \times S_R} = d_{S_L} + d_{S_R}$.
We \emph{define} $H_S(d)$ by the surface recursion and by imposing factorization of $\hat{H}_S(d)\coloneqq H_S(d)/d_s$ on disconnected surfaces.
The same stochastic process defined by \eqref{eq:probs} generate samples according to a new distribution $\mathcal{J}_{d}$, which is given by the same explicit formula \eqref{eq:explicit}.
This allows to define an interesting space of sampling distributions $\mathcal{J}_{d}$, positively parametrized by the values of the kernel $d = \{d_S \ge0, S \in \mathrm{ConnectedSurfaces}\}$. 

From this larger perspective the Hepp bound is a specific point in the space of distributions, corresponding to the choice $d_S = E_S -L_S$.
This specific point is not arbitrary: the Hepp curve integrand is given to us by the tropical limit of some physical curve integrand; here we have considered $\mathrm{Tr\ } \phi^3$ in $D=2$.
To get more accurate sampling distributions, or to evaluate amplitudes for different theories, one could certainly envision a similar strategy, where one defines a sampling distribution tailored for a desired integrand $\mathcal{I}^{\rm phys}$ by dividing the curve space in sectors, and expanding $\mathcal{I}^{\rm phys}$ in each sector up to some order.
But the viewpoint we have introduced here opens up another exciting possibility.
We can explore the space of distributions, starting from an arbitrary reference point and changing the values $d_S$ looking for a better sampling distribution. 

In order to decide in which direction to move in this parametric space of distributions we need a concrete measure of how ``good'' a sampling distribution is for integrating numerically some target physical integrand. 
It is natural for us to consider the \emph{Kullback–Leibler divergence} (KL-divergence),
\begin{align}
    D_{\rm KL}(\mathcal{J}^{\rm phys}||\mathcal{J}_d) = \int_{\mathbb{P}^{E_S-1}} \frac{\omega\ }{\mathrm{GL}(1)\times \mathrm{MCG}_S} \mathcal{J}^{\rm phys}\log \left(\frac{\mathcal{J}^{\rm phys}}{\mathcal{J}_d}\right) \ge0,
    \label{eq:KLdiv}
\end{align}
where $\mathcal{J}^{\rm phys}$ is the distribution defined by normalizing the desired physical integrand $\mathcal{I}^{\rm phys}$ by the value $A_S$.
In the following let us assume that $\mathcal{J}^{\rm phys}$ is positive\footnote{This assumption is justified only when an Euclidean region exists (and one is interested in evaluating amplitudes there). This is the case e.g. for massive $\mathrm{Tr}(\phi^3)$ and for planar massless theories. In a more realistic setting, one would have to re-introduce Feynman's $i\epsilon$, which could be done along the lines of \cite{Borinsky_2023,Hannesdottir_2022}. This would go far beyond the scope of this paper and for the time being we ignore this important subtlety.}.
Loosely speaking, the KL-divergence measure how much two probability distributions are similar to each other.
It is a non-negative function which vanishes if and only if $\mathcal{J}^{\rm phys} = \mathcal{J}_d$, therefore we are interested in minimizing it.

The reason why the KL-divergence is a natural measure for us to consider is that $D_{\rm KL}(\mathcal{I}^{\rm phys},\mathcal{J}_d)$ is \emph{convex} with respect to the parameters $d$.
In order to prove this we have to evaluate the Hessian  of \eqref{eq:KLdiv}. Since $\mathcal{J}^{\rm phys}$ does not depend on $d$ we can drop it from the logarithm. The rest can be evaluated by decomposing the Feynman fan in sectors,
\begin{align}
    \partial_d \partial_{d'} D_{\rm KL}(\mathcal{J}^{\rm phys}||\mathcal{J}_d) &= -\int_{\mathbb{P}^{E_S-1}} \frac{\omega\ }{\mathrm{GL}(1)\times \mathrm{MCG}_S} \mathcal{J}^{\rm phys}\partial_d \partial_{d'} \log\left(\frac{\mathcal{I}_d}{H_S(d)}\right) \nonumber \\
    &=\partial_d \partial_{d'} \log H_S(d)-\sum_{\mathcal{S} \in \mathrm{Sectors}(S)} \int_{\Delta_{\mathcal{S}}} d\alpha\ \mathcal{J}^{\rm phys}  \partial_d \partial_{d'}\sum_i d_{S_{(i)}} \log\left(\frac{\alpha_{i}}{\alpha_{i-1}}\right) \nonumber\\
    &= \partial_d \partial_{d'} \log H_S(d).
    \label{eq:KLconv}
\end{align}
The result then follows from the convexity of the logarithm of the partition function $H_S(d)$.

Due to its convexity, the minimization of the KL-divergence can be approached efficiently via \emph{gradient descent}. 
The derivatives of the KL-divergence are given by
\begin{align}
    \partial_d D_{\rm KL}(\mathcal{J}^{\rm phys}||\mathcal{J}_d) = -\int_{\mathbb{P}^{E_S-1}} d\mu_d\ \frac{\mathcal{J}^{\rm phys}}{\mathcal{J}_d} \partial_{d} \log \mathcal{J}_d,
    \label{eq:KLgrad}
\end{align}
where 
\begin{align}
d\mu_d = \frac{\omega}{\mathrm{GL}(1)\times \mathrm{MCG}_S}\mathcal{J}_d.
\end{align}
From the above we see that if $d\to0$ the gradient becomes large and positive due to the term coming from the partition function. Therefore the boundaries of the parametric space are repulsive for the gradient flow.

Let us make one further comment regarding the use of the KL-divergence.
Looking back at \ref{eq:mcaccuracy}, the variance $\mathrm{Var}(\mathcal{I}^{\rm phys}/\mathcal{J}{_d})$ seems a more natural measure to minimize, since it is directly related to the accuracy of the Monte Carlo estimator. However, we have no reason to expect it to be convex with respect to the parameters $d$.
On the other hand, an elementary calculation shows that its gradient is related to that of the KL-divergence, 
\begin{align}
    \partial_d\mathrm{Var}\left(\frac{\mathcal{I}^{\rm phys}}{\mathcal{J}_d}\right) \propto \int_{\mathbb{P}^{E_S-1}} d\mu_d\left(\frac{\mathcal{J}^{\rm phys}}{\mathcal{J}_d}\right)^2 \partial_d \log \mathcal{J}_d.
    \label{eq:gradDiv}
\end{align}
Compared to \eqref{eq:KLgrad}, the integrand above differs by an extra factor of $\mathcal{J}^{\rm phys}/\mathcal{J}_d$, therefore it is reasonable to expect that a small KL gradient will imply a small gradient for the variance.

We can numerically evaluate the integral in \eqref{eq:KLgrad} by generating samples distributed according to $\mathcal{J}_d$, through the stochastic process, and then evaluating the derivative $\partial_d \log \mathcal{I}_d$ in the sampled sectors using the explicit formula \eqref{eq:explicit}.
Note that this also requires calculating the derivatives of the partition function, $\partial_d \log H_S(d)$, which can be done using the recursive relation \eqref{eq:recursion}.

Let us illustrate the calculation of the gradient of the KL-divergence in a simple tree level example.
\begin{example}
    Let us consider $S_n = \{0,\{n\}\}$ and a generic recursion kernel $d_{S_n} \coloneqq d_n$. Solving the recursion, we find the partition function for the first few surfaces to be 
    \begin{align}
        H_{4} &= 2 \hat{H_3}\times\hat{H_3} = 2, \quad H_5 = 5\times\left(1\times\frac{2}{d_4}\right),\\
        H_6 &= 6\times\left(\frac{10/ d_4}{d_5}\right) + 3 \times \left(\frac{2}{d_4}\times\frac{2}{d_4}\right) .
    \end{align}
    Let us study the gradient of the KL-divergence, by computing explicitly the derivatives $\partial_{d_m} \log \mathcal{I}_{n}$.
    For $n=5$, fixing $d_3 = 1$, in any sector we have
    \begin{align}
       \left.\partial_{d_4} \log \mathcal{J}_{5}\right|_{\mathcal{S}} = \partial_{d_4} \log\frac{\alpha_2^{d_4}}{10/d_4} = \log \alpha_2 + \frac{1}{d_4}.
    \end{align}
    In any sector the tree level curve integrand for $\mathrm{Tr\ } \phi^3$ theory, at zero external momentum, restricts to $\left(1+\alpha_2\right)^{-2}$, therefore \eqref{eq:KLgrad} reads
    \begin{align}
        \partial_{d_4} D_{\rm KL}(\mathcal{J}^{\rm phys}||\mathcal{J}_d) = 10 \times \int_0^1 d\alpha_2 \frac{2}{(1+\alpha_2)^2} \left(\log (\alpha_2) + \frac{1}{d_4}\right),
    \end{align}
    which vanishes for $d_4=d_4^{\rm optimal} \sim 0.72 < d_4^{\rm Hepp}=1$.
    Simply put, we discover that the physical integrand is better approximated by the piecewise monomial distribution $\mathcal{J}=\frac{0.72}{10}\ \alpha_2^{0.72-1}$ than it is by the piecewise uniform distribution $\mathcal{J}=1/10$.
\end{example}

Having evaluated numerically the gradient, we take a small step in the space of sampling distributions choosing a new kernel $d' = d +  \epsilon\ \nabla_d D(\mathcal{J}^{\rm phys}||\mathcal{J}_d)$. We then use the surface recursion \eqref{eq:recursion} to update the stored values of the probabilities \eqref{eq:probs}.
This iterative procedure can be used to \emph{train} the stochastic process, yielding a better sampling distribution.

As a proof-of-principle of this idea, we have made the following numerical experiment.
We have initialized randomly the recursion kernel and evaluated amplitudes for a simple two-loop surface ($S= \{1,\{2\}\}$). We have measured the accuracy of the resulting Monte Carlo integration, as well as the KL-divergence and the variance.
We have then performed a few gradient descent steps and measured the accuracy of the new estimator. We have seen that after $10$ steps the accuracy, using the same number of points, increases by a digit, thus demonstrating that the method is able to find better sampling distributions even when the tropical integrand is not available.

The strategy is in principle applicable to \emph{any} curve integrand $\mathcal{I}^{\rm phys}$, opening up an interesting path to extend the applicability of the dual sampling algorithm to more general scattering amplitudes.

\section{Conclusions and Future Directions}
\label{sec:conclusions}

We end the paper by briefly summarizing the main results and by outlining a strategy to extend our methods to more general physical calculations.

The central result of this manuscript is the \emph{dual sampling algorithm} \eqref{alg:dualsampler}, that allows to generate random points on the Feynman fan distributed according to the tropical curve integrand $\mathcal{I}^{\rm tr}$ defined in Eq. \eqref{eq:trint}.
The algorithm is based on the new decomposition of curve integrals formed by the cells Eq. \eqref{eq:barycell}. It can be thought of as the barycentric dual of the triangulation of curve integrals that reproduces the Feynman diagrams expansion. The new decomposition leads to a recursive formula for the partition function of $\mathcal{I}^{\rm tr}$, as illustrated in Section \ref{sec:tropicalization}, which is the cornerstone of the dual sampling algorithm.

We have applied the dual sampling algorithm to the numerical evaluation of an infinite family of scattering amplitudes in the massive $\mathrm{Tr}(\phi)^3$ theory in $D=2$ space-time dimensions, which are free of UV and IR divergences. These are the amplitudes associated to un-punctured surfaces in the t'Hooft topological expansion \eqref{eq:surfaceordering}.
The starting point is the description \eqref{eq:curveint} of the color-ordered amplitudes as \emph{curve integrals} over the Feynman fan, which we evaluate by a Monte Carlo integration using $\mathcal{I}^{\rm tr}$ as a sampling distribution.

We provide a proof-of-principle implementation of this algorithm in a \verb|Mathematica| notebook.
We have validated our code at low loop orders by an explicit comparison with the standard expansion of amplitudes in Feynman fatgraphs, Eq. \eqref{eq:curveint0}, which we have evaluated using the public software \verb|feyntrop|\cite{Borinsky_2023,feyntrop}.
We have used the attached notebook to evaluate amplitudes with up to 10-loops, reporting the results in Table \ref{tab:results}. We find that the dual sampling algorithm is able to achieve the same accuracy as the direct Feynman approach requiring much fewer samples, and avoiding the explicit enumeration of all fatgraphs. This results in drastic computational advantages in terms of time and space requirements, which makes it possible to reach such high loop orders.

We have also proposed a strategy to both improve the performance of our method, as well as to extend its applicability to more general theories.
The idea is to consider the stochastic process implementing the sampling procedure as a parametrization for a space of sampling distributions. Within this space, we look for the  best sampling distribution to integrate a desired physical integrand.
The optimal sampling distribution can be found minimizing a convex cost function by gradient descent.
The validity of this method is demonstrated in simple examples provided in the ancillary file.
This strategy can be used to compute amplitudes beyond $\mathrm{Tr}(\phi^3)$ theory, assuming that a locally finite curve integrand is available, in parametric form, and that it can be numerically evaluated efficiently.

We believe that the ideas put forward in this paper raise a fresh perspective on the evaluation of scattering amplitudes.
Since the pioneering result of Parke and Taylor \cite{Parke:1986gb}, modern developments on scattering amplitudes have been centered around the discovery of surprisingly \emph{compact analytic expressions} for the amplitudes, and on understanding the hidden structures at their origin.
The picture emerging from this paper shifts the focus from ``space'' to ``time'', in the sense that we set ourselves the objective of finding scattering integrands that can be numerically evaluated quickly; even though they may be given by large analytical expressions.
As in previous developments, at the heart of this approach is the discovery of hidden structures --- the Feynman fan and the stochastic process on surfaces --- which render feasible calculations that would seem hopeless from a conventional point of view.
In this sense, it is particularly amusing that already in the elementary $\mathrm{Tr}(\phi^3)$ theory these structures provide such a drastic advantage over Feynman diagrams: in this theory all diagrams come with the same numerator, ``$+1$'', so that no magical cancellations can take place among them. Therefore, the total amplitude simply is as big and complicated as Feynman diagrams suggest it is.
And yet, precisely because of our shift in focus to fast numerical evaluation rather than compact analytical expressions, we have found that \emph{there are} structures that come to our rescue and --- even more intriguing --- that these are invisible from a graph-by-graph point of view. Indeed, the dual barycentric decomposition underpinning our sampling algorithm is maximally incompatible with Feynman diagrams.

In light of this new perspective, we highlight some concrete steps to make further progress as well as some other more general questions that seem interesting to consider.

{\bf UV and IR divergences.}  Scattering amplitudes suffer from ultraviolet and infrared divergences, which can be both treated in dimensional regularization.
These divergences appear in intermediate calculations only, and cancel when counter-terms and/or sufficiently inclusive observables are considered. Unfortunately, these cancellations of divergences are not enough to evaluate directly the final finite result by a Monte Carlo integration.
In order to achieve that, it is necessary to enforce the cancellations of divergences \emph{locally} at the integrand level, in the sense of \cite{Salvatori:2024nva}.
Working in parametric space, a quasi-systematic method to do so is provided by \emph{sector decomposition}\cite{Binoth:2000ps,Binoth:2003ak,Bogner_2008,Kaneko_2010,Smirnov:2021rhf}\footnote{The reason why this is not entirely systematic is that in the presence of relative signs among the coefficients of the Symanzik polynomial $\mathcal{F}_G$, one cannot guarantee that all singularities of the integrand will be correctly identified.}. It consists in dividing a Feynman simplex into smaller sectors (possibly more general than the permutohedral sectors we considered here), and use series expansions of the integrand as counter-terms in a local subtraction formula.
As it was argued at length in \cite{Salvatori:2024nva}, sector decomposition is not particularly useful if one is interested in finding analytic formulae. On the other hand, the shift in focus to quick numerical evaluation makes it an appealing strategy. The reason is that the series expansions used to construct the counter-terms \emph{commute} with each other, therefore the subtraction formula \emph{factorizes}, and can be evaluated quickly. 

{\bf Beyond $\mathrm{Tr}(\phi^3)$.} Understanding how to deal with divergences is a pre-requisite for the most interesting question, which is how to calculate amplitudes for more general theories.
The simplest extension is to let go of color, that is to move from $\mathrm{Tr}(\phi^3)$ theory to plain $\phi^3$ theory. As it was already argued in \cite{counting1}, this can be done easily within surfaceology: it suffices to consider \emph{closed curves} on surfaces.
A more interesting extension is to include spinning particles, which at the level of diagrams results in additional \emph{numerator} structures.
Numerators can be converted into parametric ones by applying Wick's theorem to the loop momentum gaussian integration (See e.g. \cite{Hannesdottir_2022,Bellon:2022qrr} for further details). Given our concern about quick evaluation, it is important to avoid the need to explicitly enumerate all the Wick's contractions, much in the same way as we avoid listing all the cuts of a graph when evaluating the Symanzik polynomials $\mathcal{U}_G$ and $\mathcal{F}_G$. 

Another important aspect to take into account is the number of numerator structures appearing in any given diagram. For concreteness, consider the case of pure Yang-Mills theory, where color ordered Feynman rules involve three possible numerators per vertex. This results in an exponential growth of the number of different numerator structures \emph{per diagram}!
This is the place where we expect the second miracle of surfaceology, which we have repeatedly referred to, to play an essential role. In \cite{scaffolding} it was proposed that the loop integrand for pure Yang-Mills theory can be extracted as a low energy limit of a stringy version of curve integrals. Surprisingly, the calculation involves very similar steps as those appearing in sector decomposition. Therefore, we speculate that it should be possible to modify our algorithm to sample over \emph{words} rather than merely over curves up to MCG\footnote{The homotopy class of a curve is identified by a word that records the way it travels across a triangulation of the surface, for further details see \cite{counting1}.}, and that as the process evolves we can extract the desired integrand step-by-step. 

{\bf Beyond tropical sampling.} 
Assuming that a fast curve integrand is available, the next step would be to study its tropicalization.
Beyond $\mathrm{Tr}(\phi^3)$ theory in $D=2$, this could be problematic, due to the appeareance of exotic Newton polytopes and pesky minus signs.
In this regard the AI-inspired idea proposed in Section \ref{sec:samplingproblem} deserves a thorough experimentation as it may provide a practical alternative to tropical sampling: one could simply start from an arbitrary sampling distribution and learn a better one by gradient descent.

In addition, the structure of this gradient flow is interesting \emph{per se}. Is there some recursive structure, where models $\mathcal{J}_{d^{\rm optimal}}$ trained for lower loop calculations can be recycled by grafting them in the stochastic processes at higher loops? Are there analogies among different physical theories, with
a model trained on some theory performing well on another? 
See \cite{hashemi2025transformersenumerativegeometry} for a recent exploration of applications of artificial intelligence to similar questions in enumerative geometry.

{\bf Integrals over Moduli spaces.} The Feynman fan can be interpreted as the tropical boundary of the \emph{Teichm\"uller} and \emph{Moduli} spaces of surfaces. Therefore, dual sampling algorithms can be uplifted to sampling over moduli spaces, allowing to evaluate integrals defined over them.
It would be interesting to leverage on this idea to further stimulate recent progresses on the numerical evaluation of string amplitudes. \cite{Eberhardt:2023xck,Baccianti:2025gll,Banerjee:2024ibt,Eberhardt:2024twy,Arkani-Hamed:2024nzc}.

{\bf{Headlight Functions.}} 
We have mentioned several times the important distinction between the first and second miracle of surfaceology, and how we have made use here only of the first.
It is natural to speculate that the second could play an important role in sampling as well. Let us give an example of how this could happen.
First, we can turn the problem of sampling over graphs into the problem of sampling over a polytope, by considering the polytope $\mathcal{P}_S$ used in the proof of the surface recursion (recall that it is triangulated by the uni-modular Feynman fan!).
The headlight functions give an efficient membership oracle to decide if a point $t$ belongs to $\mathcal{P}_S$. We can therefore sample the polytope by a random walk along the coordinate directions, which is known to thermalize in polynomial time with respect to the dimension of the polytope \cite{10.1145/102782.102783}.
To perform importance sampling, we would have to understand how to sampling non uniformly over the polytope. We leave a thorough exploration of this to the future.

\section{Acknowledgments}

We wish to thank Nima Arkani-Hamed and Hadleigh Frost for discussions and collaborations on various topics related to surfaceology and recursions.
We wish to thank Michael Borinsky for his beautiful research and for numerous conversations around the topic of Feynman integrals and tropical geometry.
We also thank Baran Hashemi and Lukas Heinrich for conversations around the applications of machine learning to particle physics.

The work of the author is part of the PositiveWorld project funded by the European Union’s Horizon 2023
research and innovation programme under the Marie Skłodowska-Curie grant agreement 101151760.
Views and opinions of the authors expressed are those of the author(s) only and do not necessarily reflect those of the European Union or the European Research Council Executive Agency. Neither the European Union nor the granting authority can be held responsible for them.
The work of G.S. is also supported in part by the DOE (Grant No. DE-SC0009988), further support was made possible by the Carl B. Feinberg cross-disciplinary program in innovation at the IAS. 

\bibliographystyle{JHEP}
\bibliography{refs}

\end{document}